\documentclass[twoside,11pt]{article}

\PassOptionsToPackage{hyphens}{url}
\usepackage{jmlr2e}
\usepackage{float}
\usepackage{booktabs}

\jmlrheading{1}{2020}{}{02/20}{N/A}{henderson2020climate}{Peter Henderson, Jieru Hu, Joshua Romoff, Emma Brunskill, Dan Jurafsky, Joelle Pineau}
\usepackage{lastpage}
\jmlrheading{21}{2020}{1-\pageref{LastPage}}{4/20; Revised
10/20}{11/20}{20-312}{Peter Henderson, Jieru Hu, Joshua Romoff, Emma Brunskill, Dan Jurafsky, Joelle Pineau}
\ShortHeadings{Towards the Systematic Reporting of the Energy and Carbon Footprints of ML}{Henderson, Hu, Romoff, Brunskill, Jurafsky, and Pineau}

\firstpageno{1}
\usepackage[utf8]{inputenc} %
\usepackage[T1]{fontenc}    %
\usepackage{booktabs}       %
\usepackage{amsfonts}       %
\usepackage{amsmath}
\usepackage{nicefrac}       %
\usepackage{microtype}      %
\usepackage{graphicx}
\usepackage{tikz}
\usepackage{tikz-uml}
\usepackage{textcomp}

\usepackage{listings}
\usepackage{xcolor}

\definecolor{codegreen}{rgb}{0,0.6,0}
\definecolor{codegray}{rgb}{0.5,0.5,0.5}
\definecolor{codepurple}{rgb}{0.58,0,0.82}
\definecolor{backcolour}{rgb}{0.95,0.95,0.92}

\lstdefinestyle{mystyle}{
    backgroundcolor=\color{backcolour},   
    commentstyle=\color{codegreen},
    keywordstyle=\color{magenta},
    numberstyle=\tiny\color{codegray},
    stringstyle=\color{codepurple},
    basicstyle=\ttfamily\footnotesize,
    breakatwhitespace=false,         
    breaklines=true,                 
    captionpos=b,                    
    keepspaces=true,                 
    numbers=left,                    
    numbersep=5pt,                  
    showspaces=false,                
    showstringspaces=false,
    showtabs=false,                  
    tabsize=2
}

\usepackage{gensymb}
\usepackage[export]{adjustbox}%
\newcommand{\coeq}{$\text{CO}_{2eq}$}
\newtheorem{thm}{Theorem}
\newtheorem{exampleinline}[thm]{Example} 
\title{Towards the Systematic Reporting of the Energy and Carbon Footprints of Machine Learning}

\author{\name Peter Henderson \email phend@cs.stanford.edu \\
       \addr Stanford University, Stanford, CA, USA
\AND \name Jieru Hu \email jieru@fb.com \\
\addr Facebook, Menlo Park, CA, USA \AND \name Joshua Romoff \email joshua.romoff@mail.mcgill.ca \\ \addr Mila, McGill University, Montreal, QC, Canada \AND \name Emma Brunskill \email ebrun@cs.stanford.edu \\
       \addr Stanford University, Stanford, CA, USA
\AND \name Dan Jurafsky \email jurafsky@stanford.edu \\
       \addr Stanford University, Stanford, CA, USA
\AND Joelle Pineau \email jpineau@cs.mcgill.ca\\
\addr Facebook AI Research, Mila, McGill University, Montreal, QC, Canada
}

\editor{David Sontag}

\begin{document}
\maketitle

\begin{abstract}%
Accurate reporting of energy and carbon usage is essential for understanding the potential climate impacts of machine learning research. 
We introduce a framework that makes this easier by providing a simple interface for tracking realtime energy consumption and carbon emissions, as well as generating standardized online appendices.
Utilizing this framework, we create a leaderboard for energy efficient reinforcement learning algorithms to incentivize responsible research in this area as an example for other areas of machine learning.
Finally, based on case studies using our framework, we propose strategies for mitigation of carbon emissions and reduction of energy consumption. 
By making accounting easier, we hope to further the sustainable development of machine learning experiments and spur more research into energy efficient algorithms.
\end{abstract}

\begin{keywords}
  energy efficiency, green computing, reinforcement learning, deep learning, climate change
\end{keywords}

\section{Introduction}
Global climate change is a scientifically well-recognized phenomenon and appears to be accelerated due to greenhouse gas (GHG) emissions such as carbon dioxide or equivalents (\coeq)~\citep{crowley2000causes,ipcc15}. The harmful health and safety impacts of global climate change are projected to ``fall disproportionately on the poor and vulnerable''~\citep{ipcc15}. 
Energy production remains a large factor in GHG emissions, contributing about $\sim 25$\% of GHG emissions in 2010~\citep{ipcc15}. 
With the compute and energy demands of many modern machine learning (ML) methods growing exponentially~\citep{aiandcompute}, ML systems have the potential to significantly contribute to carbon emissions. 
Recent work has demonstrated these potential impacts through case studies and suggested various mitigating strategies~\citep{strubell2019energy,DBLP:journals/corr/abs-1907-10597}.

Systematic and accurate measurements are needed to better estimate the broader energy and carbon footprints of ML---in both research and production settings. 
Accurate accounting of carbon and energy impacts aligns incentives with energy efficiency~\citep{DBLP:journals/corr/abs-1907-10597}, raises awareness, and drives mitigation efforts~\citep{sundar2018modeling,lariviere2016better}, among other benefits.\footnote{See Section~\ref{sec:account} for an extended discussion on the importance of accounting.}
Yet, most ML research papers do not regularly report energy or carbon emissions metrics.\footnote{See Section~\ref{sec:background} and Appendix~\ref{app:neurips} for more information.}

We hypothesize that part of the reason that much research does not report energy and carbon metrics is due to the complexities of collecting them. Collecting carbon emission metrics requires understanding emissions from energy grids, recording power outputs from GPUs and CPUs, and navigating among different tools to accomplish these tasks.
To reduce this overhead, we present \emph{experiment-impact-tracker}\footnote{\url{https://github.com/Breakend/experiment-impact-tracker}}---a lightweight framework for consistent, easy, and more accurate reporting of energy, compute, and carbon impacts of ML systems.

In Section~\ref{sec:tool}, we introduce the design and capabilities of our framework and the issues with accounting we aim to solve with this new framework.
Section~\ref{sec:difficulties} expands on the challenges of using existing accounting methods and discusses our learnings from analyzing experiments with \emph{experiment-impact-tracker}.
For example, in an empirical case study on image classification algorithms, we demonstrate that floating point operations (FPOs), a common measure of efficiency, are often uncorrelated with energy consumption with energy metrics gathered by \emph{experiment-impact-tracker}.

In Section~\ref{sec:mitigation}, we focus on recommendations for promoting energy-efficient research and mitigation strategies for carbon emissions.
Using our framework, we present a \emph{Reinforcement Learning Energy Leaderboard} in Section~\ref{sec:rl} to encourage development of energy efficient algorithms.
We also present a case study in machine translation to show how regional energy grid differences can result in large variations in \coeq  emissions. Emissions can be reduced by up to 30x just by running experiments in locations powered by more renewable energy sources (Section~\ref{sec:regional_carbon}).
Finally, we suggest systemic and immediate changes based on our findings:
\begin{itemize}
    \item incentivizing energy-efficient research through leaderboards (Section~\ref{sec:leaderboards})
    \item running experiments in carbon-friendly regions (Section~\ref{sec:regional_carbon})
    \item  reducing overheads for utilizing efficient algorithms and resources (Section~\ref{sec:ease-of-use})
    \item considering energy-performance trade-offs before deploying energy hungry models (Section~\ref{sec:cost-benefit})
    \item selecting efficient test environment especially in RL (Section~\ref{sec:efficiency-test-env})
    \item ensuring reproducibility to reduce energy consumption from replication difficulties (Section~\ref{sec:reproducibility})
    \item consistently reporting energy and carbon metrics (Section~\ref{sec:standard})
\end{itemize}

\section{Related Work}
\label{sec:related}

Estimating GHG emissions and their downstream consequences is important for setting regulatory standards~\citep{sheet2013social} and encouraging self-regulation~\citep{byerly2018nudging}. In particular, these estimates are used to set carbon emissions reduction targets and in turn set carbon prices for taxes or emissions trading systems.\footnote{An emissions trading system is a cap on total allowed carbon emissions for a company with permits issued. When a company emits a certain amount of carbon, they trade in a permit, creating a market for emissions permits. This is a market-based approach to incentivize emission reductions. See \citet{ramstein2019state} for a description of such carbon pricing efforts across different countries.}
A large body of work has examined modeling and accounting of carbon emissions\footnote{See also assorted examinations on carbon accounting, standardized reporting, and policy recommendations \citep{stechemesser2012carbon,dayarathna2015data,ipcc15,ajani2013comprehensive,bellassen2015accounting,andrew2011accounting,tang2018climate,cotter2011standardized,tol2011social,sheet2013social,countrylevelsocialcostofcarbon}.} at different levels of granularity: at the global scale~\citep{ipcc15}; using country-specific estimates~\citep{countrylevelsocialcostofcarbon}; targeting a particular industrial sector like Information and Communication Technologies, for example, modeled by \citet{malmodin2013future}; or even targeting a particular application like bitcoin mining, for example, modeled by~\citet{mora2018bitcoin}.

At the application level, some work has already modeled carbon impacts specifically in computationally intensive settings like bitcoin mining~\citep{krause2018quantification,stoll2019carbon,zade2019bitcoin,mora2018bitcoin}. Such application-specific efforts are important for prioritizing emissions mitigation strategies: without understanding projected impacts, policy decisions could focus on ineffective regulation. However, with large amounts of heterogeneity and endogeneity in the underlying data, it can be difficult to model all aspects of an application's usage. For example, one study suggested that ``bitcoin emissions alone could push global warming above 2 \degree C''~\citep{mora2018bitcoin}. But \citet{masanet2019implausible}, \citet{houy2019rational}, and others, criticized the underlying modeling assumptions which led to such large estimates of carbon emissions. This shows that it is vital that these models provide accurate measurements if they are to be used for informed decision making.

With ML models getting more computationally intensive~\citep{aiandcompute}, we want to better understand how machine learning in research and industry impacts climate change.
However, estimating aggregate climate change impacts of ML research and applications would require many assumptions due to a current lack of reporting and accounting.
Instead, we aim to emphasize and aid systematic reporting strategies such that accurate field-wide estimates can be conducted in the future.

Some recent work specifically investigates climate impacts of machine learning research. \citet{strubell2019energy} demonstrate the issue of carbon and energy impacts of large NLP models by evaluating estimated power usage and carbon emissions for a set of case studies. The authors suggest that: ``authors should report training time and sensitivity to hyperparameters'', ``academic researchers need equitable access to computation resources'', and ``researchers should prioritize computationally efficient hardware and algorithms''. 
\citet{DBLP:journals/corr/abs-1907-10597} provide similar proposals, suggesting floating point operations (FPOs) as a guiding efficiency metric. 
\citet{lacoste2019quantifying} recently provided a website for estimating carbon emissions based on GPU type, experiment length, and cloud provider. 
In Section~\ref{sec:difficulties}, we discuss how while the estimation methods of these works provide some understanding of carbon and energy impacts, nuances in the estimation methods may make them inaccurate---particularly in experiments which utilize combined CPU and GPU workloads heavily.
We build a framework aiming to provide  more accurate and easier systematic reporting of carbon and energy footprints. 
We also provide additional mitigation and reporting strategies---beyond those discussed by these prior works---to emphasize how both companies and research labs can be more carbon and energy efficient. 

It is worth noting that prior work has also examined the carbon impacts of research in other fields, focusing mostly on emissions from conference travel~\citep{spinellis2013carbon,astudillo2018estimating,psychimpact}. We provide a brief discussion on ML-related conference travel in Appendix~\ref{sec:conference_travel}, but will focus mainly on accurate accounting of energy and carbon footprints of ML compute.

\section{Background}
\label{sec:background}
We briefly provide a primer on energy and carbon accounting, which form the basis of our proposed framework for measuring and reporting the ecological footprint of ML research.

\subsection{Energy Accounting}
 Energy accounting is fairly straightforward. The energy consumption of a system can be measured in Joules (J) or Watt-hours (Wh),\footnote{One Watt is a unit of power---equivalent to one Joule per second.} representing the amount of energy needed to power the system. Life-cycle accounting might also consider the energy required to manufacture components of the system---for example, the production of GPUs or CPUs~\citep{jones2013green}. However, we largely ignore life-cycle aspects of energy accounting due to the difficulties in attributing manufacturing impacts on a per-experiment basis. Measuring data-center energy impacts also contain several layers, focusing on hardware-centric and software-centric analyses. Many parts contribute to the power consumption of any computational system. \citet{dayarathna2015data} survey energy consumption components of a data center and their relative consumption: cooling (50\%), lighting (3\%), power conversion (11\%), network hardware (10\%), and server/storage (26\%). 
 
 The server and storage component can further be broken down into contributions from DRAM, CPUs, among other compute components. Accurate accounting for all of these components requires complex modeling and varies depending on workload. In particular, the efficiency of the hardware varies with utilization---often most efficient near maximum utilization---making utilization an important factor in optimization (particularly in large cloud compute systems)~\cite{8519941}. Since we aim to provide a framework at the per-experiment software level, we only account for aspects of energy consumption which expose interfaces for energy metrics (giving us real-time energy usage and compensating for such workload differences). For the purpose of our work, this is constrained to DRAM, CPUs, and GPUs. To account for all other components, we rely on a power usage effectiveness (PUE) factor \citep{strubell2019energy}. This factor rescales the available power metrics by an average projected overhead of other components. With more available software interfaces, more robust modeling can be performed as reviewed by \citet{dayarathna2015data}. 

\subsection{Carbon Accounting}

Carbon accounting can be all-expansive, so we focus on a narrow definition provided by \citet{stechemesser2012carbon}: ``carbon accounting at the project scale can be defined as the measuring and non-monetary valuation of carbon and GHG emissions and offsetting from projects, and the monetary assessment of these emissions with offset credits to inform project-owners and investors but also to establish standardized methodologies.''
Carbon and GHG emissions are typically measured in some form close to units \coeq. This is the amount of carbon---and other GHG converted to carbon amounts---released into the atmosphere as a result of the project. Carbon offsetting is the amount of carbon emissions saved as a result of the project. For example, a company may purchase renewable energy in excess of the energy required for their project to offset for the carbon emissions they contributed. 
Since our goal is to inform and assess carbon emissions of machine learning systems, we ignore carbon offsetting.
Typical carbon offsetting involves the use of Power Purchase Agreements (PPAs) or other similar agreements which may not reflect the current carbon make-up of the power draw (as they may account for future clean energy).\footnote{See discussion in Appendix~\ref{app:carbon_discussion} for further information.} Since carbon effects contribute to feedback loops, cutting emissions now will improve the likelihood of preventing further emissions.\footnote{See, e.g., \url{https://www.esrl.noaa.gov/gmd/outreach/info_activities/pdfs/TBI_understanding_feedback_loops.pdf}}.
We also do not consider carbon accounting in the financial sense, but do provide metrics on monetary impacts through the social cost of carbon (SC-CO2). The \citet{sheet2013social} uses this metric when developing administrative rules and regulations. According to the EPA, ``The SC-CO2 is a measure, in dollars, of the long-term damage done by a ton of carbon dioxide (CO2) emissions in a given year.  This dollar figure also represents the value of damages avoided for a small emission reduction (i.e., the benefit of a CO2 reduction).'' We rely on the per-country social cost of carbon developed by \citet{countrylevelsocialcostofcarbon}, which accounts for different risk profiles of country-level policies and GDP growth in their estimates of SC-CO2. 

Carbon emissions from a project can also consider life-cycle emissions (for example, manufacturing of CPUs may emit carbon as part of the process). We do not consider these aspects of emissions. We instead, consider only carbon emissions from energy consumption. A given energy grid powering an experiment will have a carbon intensity: the grams of \coeq~emitted per kWh of energy used. This carbon intensity is determined based on the energy sources supplying the grid. Each energy source has its own carbon intensity accounted for through a full life-cycle analysis~\citep{icc2015}. For example, coal power has a median carbon intensity of 820 g\coeq / kWh, while hydroelectricity has a mean carbon intensity of 24 g\coeq / kWh. The life-cycle emissions of energy source take into account not just emissions from production, but from waste disposal as well. For example, nuclear energy waste disposal has some carbon emissions associated that would be taken into account in a life-cycle carbon intensity metric~\citep{ipcc15}. Carbon emissions for a compute system can be estimated by understanding the carbon intensity of the local energy grid and the energy consumption of the system. Similar analyses have been done for bitcoin~\citep{krause2018quantification}. These analyses, however, attempt to extrapolate impacts of bitcoin mining in general, while in this work we attempt to examine machine learning impacts on a per-experiment basis.

\subsection{Current State of Reporting in Machine Learning Research}

We briefly examine the current state of accounting in the machine learning literature and review commonly reported computational metrics. Here we look at a non-exhaustive list of reported metrics from papers we surveyed and group them into different categories:

\begin{itemize}
   \item{Energy}
   \begin{itemize}
       \item Energy in Joules~\citep{assran2019gossip}  
       \item Power consumption in Watts~\citep{canziani2016analysis}
   \end{itemize}
   \item{Compute}
   \begin{itemize}
       \item PFLOPs-hr~\citep{aiandcompute}, the floating point operations per second needed to run the experiment in one hour
       \item Floating Point Operations (FPOs) or Multiply-Additions (Madds), typically reported as the computations required to perform one forward pass through a neural network~\citep{howard2017mobilenets,sandler2018mobilenetv2,DBLP:journals/corr/abs-1907-10597}
       \item The number of parameters defined by a neural network (often reported together with FPOs)~\citep{howard2017mobilenets,sandler2018mobilenetv2}
       \item GPU/CPU utilization as a percentage~\citep{assran2019gossip,dalton2019gpuaccelerated}
       \item GPU-hours or CPU-hours, the processor cycles utilized (or in the case of the GPU percentage utilized), times the runtime~\citep{soboczenski2018bayesian}
   \end{itemize}
   \item{Runtime}
   \begin{itemize}
       \item Inference time, the time it takes to run one forward pass through a neural network,~\citep{jeon2018constructing,qin2018diagonalwise}
       \item Wall clock training time, the total time it takes to train a network~\citep{assran2019gossip,dalton2019gpuaccelerated}.
       \item Hardware and time together (e.g., 8 v100 GPUs for 5 days)~\citep{krizhevsky2012imagenet,ott2018scaling,gehring2017convolutional}
   \end{itemize}
   \item{Carbon Emissions}
   \begin{itemize}
       \item US-average carbon emissions  \citep{strubell2019energy}
   \end{itemize}   
\end{itemize}

\begin{exampleinline}
To get a rough estimate of the prevalence of these metrics, we randomly sampled 100 NeurIPS papers from the 2019 proceedings. In addition to the metrics above, we also investigate whether hardware information was reported (important for extrapolating energy and carbon information with partial information).
Of these papers, we found 1 measured energy in some way, 45 measured runtime in some way, 46 provided the hardware used, 17 provided some measure of computational complexity (e.g., compute-time, FPOs, parameters), and 0 provided carbon metrics. See Appendix~\ref{app:neurips} for more details on methodology. 
\end{exampleinline}

Some of these metrics, when combined, can also be used to roughly estimate energy or carbon metrics. For example, the experiment time (h) can be multiplied by the thermal design power (TDP) of the GPUs used (W)\footnote{This is a rough estimate of the maximum operating capacity of a GPU.}. This results in a Watt-hour energy metric. This can then be multiplied by the carbon intensity of the local energy grid to assess the amount of \coeq emitted. This method of estimation omits CPU usage and assumes a 100\% GPU utilization. Alternatively, \citet{aiandcompute} use a utilization factor of 33\% for GPUs. 
Similarly, the PFLOPs-hr metric can by multiplied by TDP (Watts) and divided by the maximum computational throughput of the GPU (in PFLOPs). This once again provides a Watt-hour energy metric. This, however, makes assumptions based on maximum efficiency of a GPU and disregards variations in optimizations made by underlying frameworks (e.g., Tensorflow versus Pytorch; AMD versus NVIDIA drivers).  

As we will demonstrate using our framework (see Section~\ref{sec:bad_estimates}), the assumptions of these estimation methods lead to significant inaccuracies. 
However, aggregating all necessary accounting information is not straightforward or easy; it requires finding compatible tools, handling nuances on shared machines, among other challenges. 

It is worth noting that some metrics focus on the computational requirements of training (which require additional resources to compute gradients and backpropagate, in the case of neural networks) versus the computational requirements of inference. The former is often more energy and carbon intensive in machine learning research, while the later is more intensive in production systems (the cost of training is insignificant when compared to the lifetime costs of running inference millions of times per day, every day). We will remain largely agnostic to this differentiation until some discussions in Sections~\ref{sec:regional_carbon} and~\ref{sec:cost-benefit}.

\section{A New Framework for Tracking Machine Learning Impacts}
\label{sec:tool}

\subsection{Motivation}
\label{sec:account}
The goal of our \emph{experiment-impact-tracker} framework is to provide an easy to deploy, reproducible, and quickly understood mechanism for all machine learning papers to report carbon impact summaries, along with additional appendices showing detailed energy, carbon, and compute metrics.

\begin{exampleinline}
A carbon impact summary generated by our framework can be found at the end of this paper in the Carbon Impact Statement section. In brief, the experiments in our paper contributed 8.021 kg of $\text{CO}_{2eq}$ to the atmosphere and used 24.344 kWh of electricity, having a USA-specific social cost of carbon of \$0.38 (\$0.00, \$0.95)~\citep{countrylevelsocialcostofcarbon}.
\end{exampleinline}

Such statements and informational reporting are important for, among other reasons, awareness, aligning incentives, and enabling accurate cost-benefit analyses.

\textbf{Awareness: } Informational labels and awareness campaigns have been shown to be effective drivers of eco-friendly behaviors (depending on the context)~\citep{banerjee2003eco,sundar2018modeling,newell2014nudging,byerly2018nudging}.
Without consistent and accurate accounting, many researchers will simply be unaware of the impacts their models might have and will not pursue mitigating strategies. Consistent reporting also may provide social incentives to reduce carbon impacts in research communities.

\textbf{Aligning Incentives: } While current reporting often focuses solely on performance metrics (accuracy in classification, perplexity in language modeling, average return in reinforcement learning, etc), standardized reporting of energy in addition to these metrics aligns incentives towards energy efficient models in research output~\citep{DBLP:journals/corr/abs-1907-10597}. Those who accurately report carbon emissions may have more incentive to reduce their carbon footprint. This may also drive traffic to low-emission regions, spurring construction of more carbon-friendly data centers.\footnote{See discussion in Section~\ref{sec:regional_carbon} on regional carbon emission differences. See discussion by \citet{lariviere2016better} on how more accurate carbon accounting can result in reduced carbon emissions.}

\textbf{Cost-Benefit Analysis and Meta-Analysis: } 
Cost-benefit analyses can be conducted with accurate energy metrics reporting, but are impossible without it.
For example, the estimated generated revenue of a model can be weighed against the cost of electricity. In the case of models suggested by \citet{rolnick2019tackling}, the carbon emissions saved by a model can be weighed against the emissions generated by the model. 
Consistent reporting also opens the possibility for performing meta-analyses on energy and carbon impacts~\citep{henderson2018distilling}. Larger extrapolations to field-wide impacts of research conferences can also be assessed with more frequent reporting.

\subsection{Design Considerations}

We consider five main principles when designing the framework for systematic reporting: usability, interpretability, extensibility, reproducibility, and fault tolerance.

\textbf{Usability: } Perceived ease-of-use can be an important factor in adoption of new technologies and methods~\citep{gefen2000relative}. Since gathering key energy ($kWh$) and carbon (\coeq) metrics requires specific knowledge about---and aggregation of---different sources of information, there may be a barrier to the ease-of-use in the current status quo. As a result, a core design consideration in developing tools for these metrics is usability, or ease-of-use. We accomplish this by abstracting away and distilling required knowledge of information sources, keeping amount of required action from the user to a minimum. %

\textbf{Interpretability: } Along with ease-of-use, a key factor in adoption is perceived usefulness~\citep{gefen2000relative}. Since we wish for the reporting of carbon and energy metrics to become widespread, we consider perceived usefulness through interpretability. We aim to make reporting tools within the framework useful through simple generation of graphs and web pages from metrics for easy interpretation. We also provide a mechanism to generate a carbon impact statement with the social cost of carbon. This dollar amount represents the projected damage from the experiment's carbon emissions and helps ground results in values that may be more interpretable. As seen in our own statement at the end of this work, we also provide the carbon impact and energy usage directly.

\textbf{Extensibility:} We design the framework in a modular fashion to handle evolving driver support (see Section~\ref{sec:difficulties}) and new metrics. To improve the accuracy and accessibility of the framework, the ML community can add new metrics, carbon intensity information, and other capabilities easily. For each metric, a central data router stores a description, the function which gathers metric data, and a list of compatibility checks (e.g., the metric can only be gathered on a Linux system). New metrics can be added to this router.\footnote{See \url{https://breakend.github.io/experiment-impact-tracker/contributing_new_metric.html}} Similarly, new carbon region and electricity grid information can be added as needed to similar centralized locations.\footnote{See \url{https://breakend.github.io/experiment-impact-tracker/contributing_carbon_region.html}.}

\textbf{Reproducibility: } Running an algorithm on different sets of hardware has been shown to affect the reproducibility of algorithmic results~\citep{gundersen2018state,sukhoy2019eliminating}. Our framework aides in automating reproducibility by logging additional metrics like hardware information, Python package versions, etc. These metrics can help future work assess statistically significant differences in model energy requirements by accounting for controlled and random variates~\citep{boquet2019decovac}.

\textbf{Fault tolerance:} Mistakes in software are inevitable---as is discussed in \cite{sidor2017openai}. We try to log all \emph{raw} information so that accounting can be recreated and updated based on new information. We also log the version number of the tool itself, to ensure future comparisons do not mismatch information between versions that may have changed.

\subsection{Proposed Framework} 

The \emph{experiment-impact-tracker} requires a simple code change to automatically gather available metrics and a script to generate online appendices for reporting the data. Currently, on compatible systems, we gather: 
\begin{itemize}
    \item all python packages and version numbers
    \item CPU and GPU hardware information
    \item experiment start and end-times
    \item the version of the \emph{experiment-impact-tracker} framework used
    \item the energy grid region the experiment is being run in (based on IP address)
    \item the average carbon intensity in the energy grid region
    \item CPU- and GPU-package power draw
    \item per-process utilization of CPUs and GPUs
    \item GPU performance states
    \item memory usage
    \item the realtime CPU frequency (in Hz)
    \item realtime carbon intensity (only supported in CA right now)
    \item disk write speed
\end{itemize}

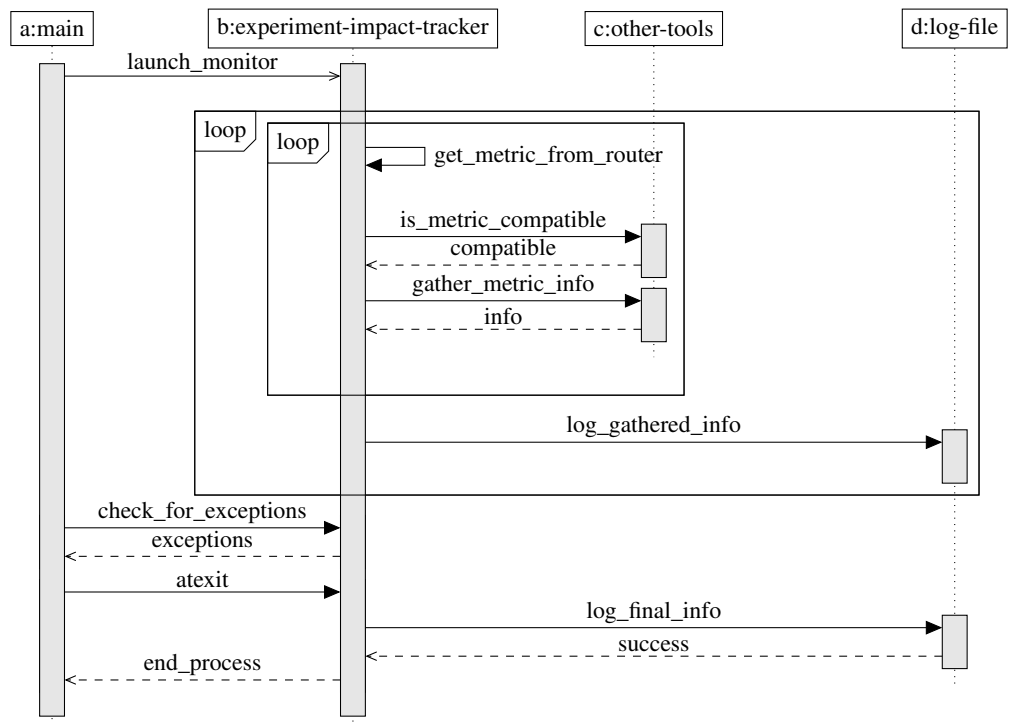
\begin{figure}[!htbp]
    \centering
    \begin{tikzpicture}
 \tikzumlset{fill object = white, fill call = gray!20} 
\begin{umlseqdiag}
\umlobject[class=main]{a}
\umlobject[class=experiment-impact-tracker]{b}
\umlobject[class=other-tools]{c}
\umlobject[class=log-file]{d}

\begin{umlcall}[op=launch\_monitor,type=asynchron]{a}{b} 
\begin{umlfragment}[type=loop]
\begin{umlfragment}[type=loop]
\begin{umlcallself}[op=get\_metric\_from\_router]{b} 
\begin{umlcall}[op=is\_metric\_compatible, return=compatible]{b}{c} 
\end{umlcall}
\begin{umlcall}[op=gather\_metric\_info, return=info]{b}{c} 
\end{umlcall}
\end{umlcallself}
\end{umlfragment}

\begin{umlcall}[op=log\_gathered\_info]{b}{d} 
\end{umlcall}
\end{umlfragment}

\begin{umlcall}[dt=35,op=check\_for\_exceptions, type=synchron,return=exceptions]{a}{b} 
\end{umlcall}

\begin{umlcall}[op=atexit, type=synchron, return=end\_process]{a}{b} 
\begin{umlcall}[op=log\_final\_info, type=synchron, return=success]{b}{d}

\end{umlcall}
\end{umlcall}
\end{umlcall}

\end{umlseqdiag}
\end{tikzpicture}
    \caption{A diagram demonstrating how the released version of the tool works. The main process launches a monitoring thread which iterates over a list of metrics associated with function calls to other tools. For example, if available, we call Intel RAPL to collect CPU power draw or query \url{caiso.org} to get realtime carbon intensity data for California. Once all the data that is compatible with the current system is gathered, it is logged to a standardized log file and the process repeats. The main thread may check in on this thread for exceptions, but the thread will not interrupt the main process. Once the main thread exits, an \emph{atexit} hook (which is called whenever the main process exits, either successfully or through an exception) gathers the final information (such as the time the experiment ended), logs it, and then ends both the monitor and main process.}
    \label{fig:process}
\end{figure}

The code change required for immediate logging of metrics can be seen in Listing 1. In the background, the framework launches a thread which polls system supported tools. For example, the thread polls \emph{psutil}~\citep{rodola2016psutil} for measuring CPU utilization. All of these metrics are logged in parallel with the main machine learning process as described in Figure~\ref{fig:process}. A script\footnote{\url{https://github.com/Breakend/experiment-impact-tracker/blob/master/scripts/create-compute-appendix}} is provided to generate an HTML web page showing graphs and tables for all these metrics, meant to serve as an online appendix for research papers.\footnote{Appendices generated by our framework for Figure~\ref{fig:transformer} and Figure~\ref{fig:imagenet} are available at: \url{https://breakend.github.io/ClimateChangeFromMachineLearningResearch/measuring_and_mitigating_energy_and_carbon_footprints_in_machine_learning/}. Experiments in Figure~\ref{fig:rl_average} are available at \url{https://breakend.github.io/RL-Energy-Leaderboard/reinforcement_learning_energy_leaderboard/index.html}.} Results in the generated appendix can be aggregated across multiple experiments to show averages along with standard error as recommended in prior work~\citep{henderson2018deep,Colas2018HowMR,Reimers2017ReportingSD}.
\lstset{style=mystyle}
\begin{lstlisting}[language=Python,caption={Simple code addition required to log experiment details via our framework.},captionpos=b]
from experiment_impact_tracker.compute_tracker import ImpactTracker
tracker = ImpactTracker(<your log directory here>)
tracker.launch_impact_monitor()
\end{lstlisting}

\subsubsection{Tracking Energy Consumption} 
\label{sec:per-process}

Different hardware vendors provide different tooling for tracking energy consumption. Our framework hides these complications from users. We currently use Intel's RAPL tool with the powercap interface~\citep{david2010rapl} or Intel's PowerGadget Tool\footnote{\href{https://software.intel.com/content/www/us/en/develop/articles/intel-power-gadget.html}{https://software.intel.com/content/www/us/en/develop/articles/intel-power-gadget.html}} (depending on availability) to gather CPU/DRAM power draw and Nvidia's \emph{nvidia-smi}\footnote{\href{https://developer.nvidia.com/nvidia-system-management-interface}{https://developer.nvidia.com/nvidia-system-management-interface}} for GPU power draw. We use \emph{psutil} for gathering per-process CPU utilization and \emph{nvidia-smi} for per-process GPU utilization. 
We found that on a shared machine---as when running a job on Slurm---using Intel's RAPL would provide energy metrics for the entire machine (including other jobs running on the worker). If two experiments were launched with Slurm to the same worker, using measurements from RAPL without corrections would double count energy usage from the CPU. 

As a result, we assign energy credits on a per-process basis (though we log system-wide information as well). We track the parent process, and any children spawned. Power credits are provided based on relative usage of system resources. If a process uses 25\% of the CPU (relative to the entire system's usage), we will credit the process with 25\% of the CPU-based power draw. This ensures that any non-experiment-related background processes--- software updates, weekly jobs, or multiple experiments on the same machine---will not be taken into account during training. 

We calculate total energy as:

\begin{equation}
    e_\text{total} = \text{PUE} \sum_p (\text{p}_\text{dram} e_\text{dram}  + \text{p}_\text{cpu} e_\text{cpu}+ \text{p}_\text{gpu} e_{\text{gpu}}), 
\end{equation} 

where $p_\text{resource}$ are the percentages of each system resource used by the attributable processes relative to the total in-use resources and $e_\text{resource}$ is the energy usage of that resource. This is the per-process equivalent of the method which \citet{strubell2019energy} use. We assume the same constant power usage effectiveness (PUE) as \citet{strubell2019energy} to be the framework's default PUE. This value compensates for excess energy from cooling or heating the data-center. Users can customize the PUE value when using the framework if needed.

\subsubsection{Carbon Accounting} 
\begin{figure}[!htbp]
    \centering
    \includegraphics[width=.9\textwidth]{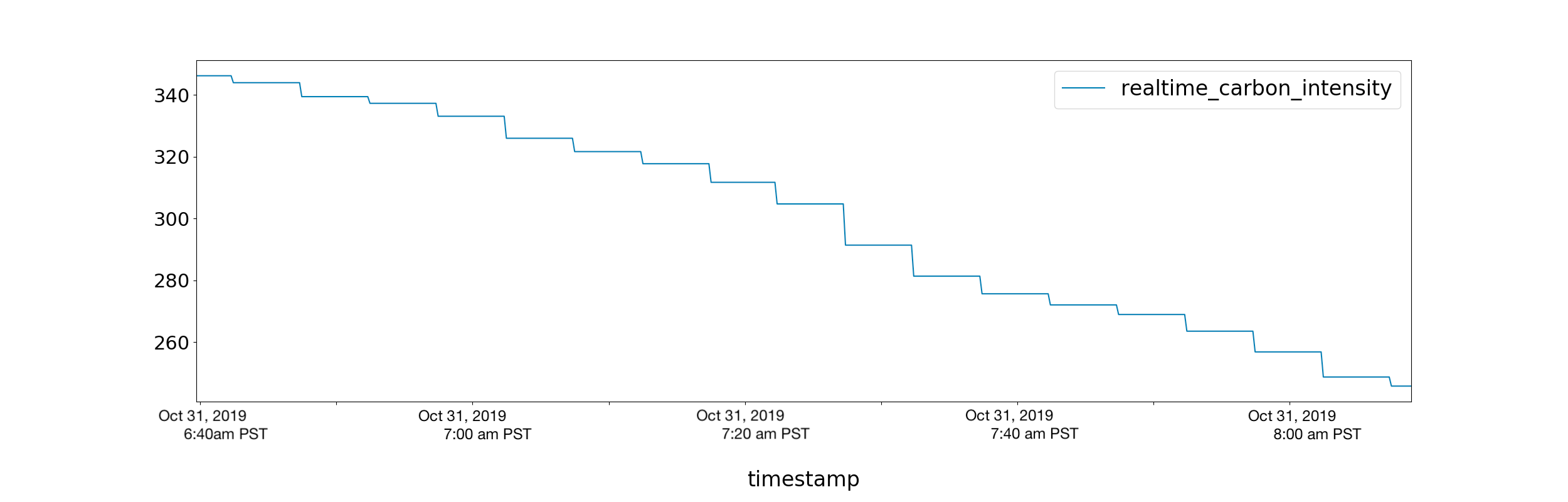}
    \caption{Realtime carbon intensity (g\coeq/kWh) collected during one experiment using our framework. As the experiment continued, the sun rose in California, and with it the carbon intensity decreased.}
    \label{fig:intensity}
\end{figure}

For calculating carbon emissions, we use the power estimate from the previous section in kilowatt-hours (kWh) and multiply it by the carbon intensity of the local energy grid (g \coeq / kWh).
To gather carbon intensity metrics for energy grids, we build on the open-source portions of \url{https://www.electricitymap.org} and define regions based on map-based geometries, using the smallest bounding region for a given location as the carbon intensity estimate of choice. For example, for an experiment run in San Francisco, if the average carbon intensity is available for both the USA and California, the latter will be used. 
We estimate the region the experiment is conducted in based on the machine's IP address.
Carbon intensities are gathered from the average fallback values provided in the \url{https://www.electricitymap.org} code where available and supplemented with additional metrics from various governmental or corporate reports. 
We note that \url{electricitymap.org} estimates are based on a closed-source system and uses the methodology described by \citet{tranberg2019real}.
All estimates from \url{electricitymap.org} are of the regional supply, rather than production (accounting for imports from other regions).
Since \url{https://caiso.com} provides realtime intensities including imports for free, for experiments run in California, we also provide realtime carbon intensity information. We do this by polling \url{https://caiso.com} for the current intensity of the California energy grid every five minutes. This helps gather even more accurate estimates of carbon emissions to account for daily shifts in supply. For example, experiments run in California during the day time use roughly $\frac{2}{3}$ of night-time experiments. This is because much of California's renewable energy comes from solar plants. Figure~\ref{fig:intensity} is an automatically generated graph showing this phenomenon from an experiment using our framework.
We hope that as users find more accurate realtime or average measurements of regional supply-based carbon intensities, they will add them to the tool for even more accurate measurements in the future.

\section{The Importance and Challenges of Accounting: Why a New Framework?}
\label{sec:difficulties}

\subsection{FPOs Can Be Misleading} 
\label{sec:flops}

Floating Point Operations (FPOs) are the de facto standard for reporting ``efficiency'' of a deep learning model~\citep{DBLP:journals/corr/abs-1907-10597}, and intuitively they should be correlated with energy efficiency---after all, fewer operations should result in faster and more energy efficient processing. However, this is not always the case.

Previously, \citet{jeon2018constructing} demonstrated mechanisms for constructing networks with larger FPOs, but lower inference time---discussing the ``Trap of FLOPs''.
Similarly, \citet{qin2018diagonalwise} show how Depthwise 3x3 Convolutions comprised just 3.06\% of an example network's Multiply-Add operations, while utilizing 82.86\% of the total training time in the FPO-efficient MobileNet architecture~\cite{howard2017mobilenets}. 
Underlying optimizations at the firmware, deep learning framework, memory, or even hardware level can change energy efficiency and run-time.
This discrepancy has led to Github Issues where users expect efficiency gains from FPO-efficient operations, but do not observe them.\footnote{See for example:  \href{https://web.archive.org/web/20191226202025/https://github.com/tensorflow/tensorflow/issues/12132}{https://github.com/tensorflow/tensorflow/issues/12132} and \href{https://web.archive.org/web/20191226202050/https://github.com/tensorflow/tensorflow/issues/12940}{https://github.com/tensorflow/tensorflow/issues/12940}}
This has also been observed by \citet{googlecvpr} and \citet{energysysml}.

\begin{figure}[!htbp]
    \centering
    \includegraphics[width=.49\textwidth]{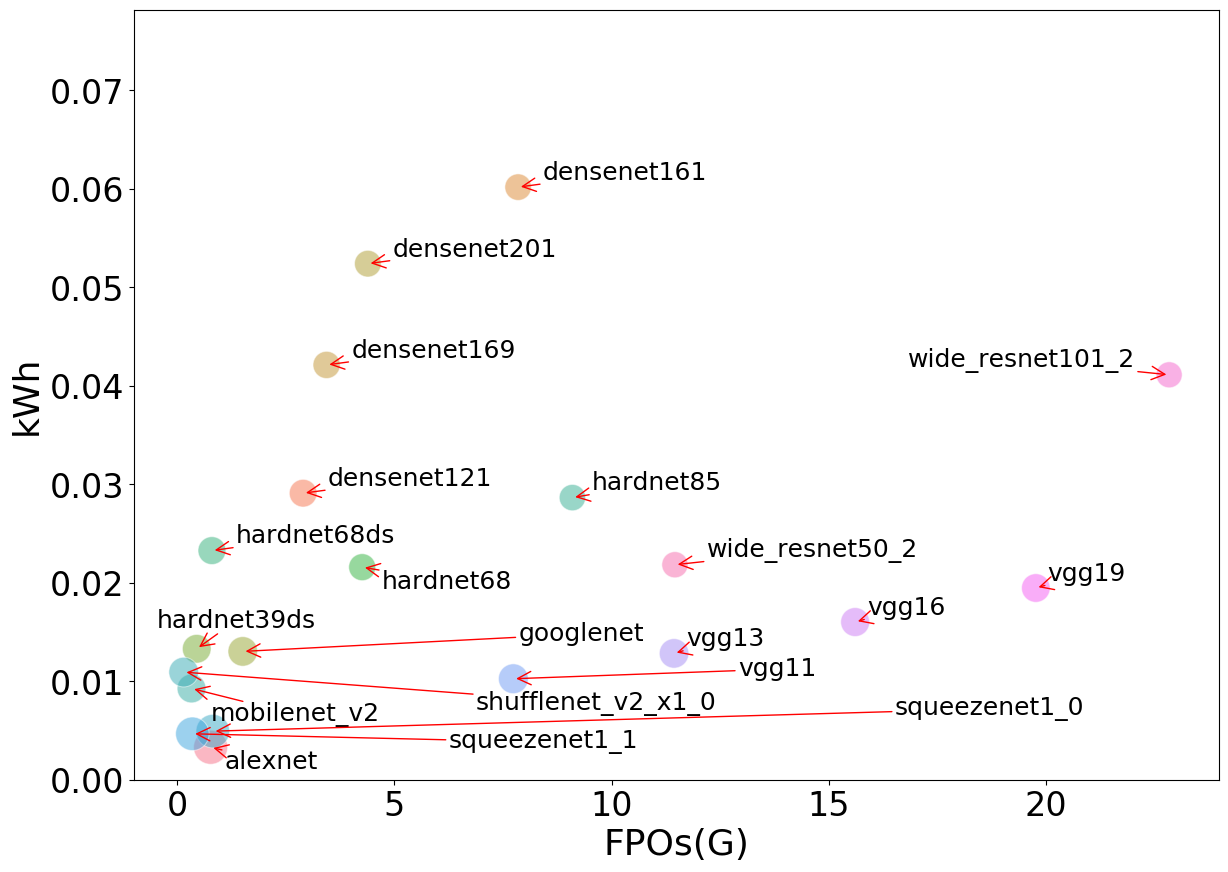}
    \includegraphics[width=.49\textwidth]{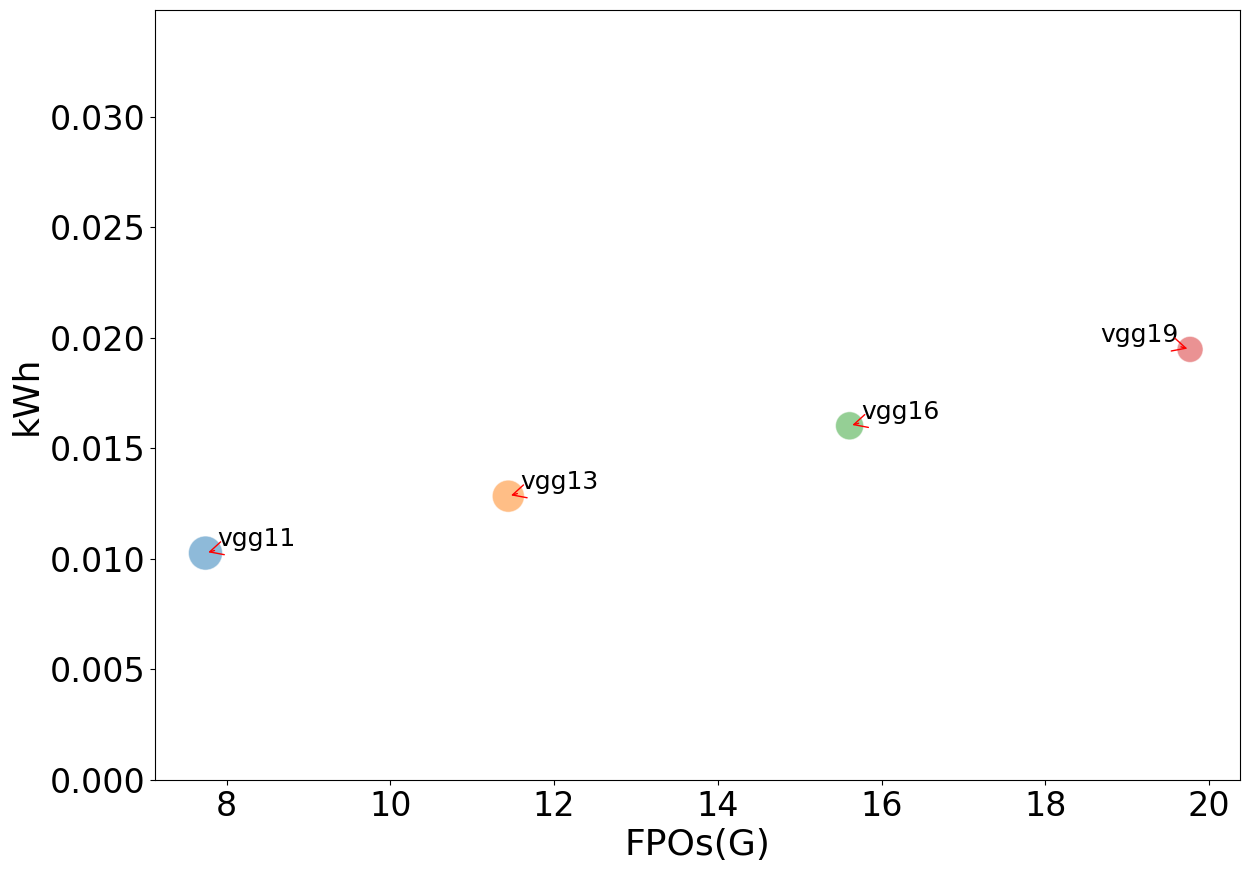}
    \caption{We run 50,000 rounds of inference on a single sampled image through pre-trained image classification models and record kWh, experiment time, FPOs, and number of parameters (repeating 4 times on different random seeds). References for models, code, and expanded experiment details can be found in Appendix~\ref{app:imagenet}. We run a similar analysis to \citet{canziani2016analysis} and find (left) that FPOs are not strongly correlated with energy consumption ($R^2=0.083$, Pearson $0.289$) nor with time ($R^2 = 0.005$, Pearson $-0.074$) when measured across different architectures. However, within an architecture (right) correlations are much stronger. Only considering different versions of VGG, FPOs are strongly correlated with energy ($R^2=.999$, Pearson $1.0$) and time ($R^2=.998$, Pearson $.999$). Comparing parameters against energy yields similar results (see Appendix~\ref{app:imagenet} for these results and plots against experiment runtime).}
    \label{fig:imagenet}
\end{figure}

\begin{exampleinline}
To investigate this empirically, we repeatedly run inference through pre-trained image classification models and measure FPOs, parameters, energy usage, and experiment length using the \emph{experiment-impact-tracker} framework. As described in Figure~\ref{fig:imagenet}, we find little correlation between FPOs and energy usage or experiment runtime when comparing across different neural network architectures. 
However, within an architecture---relying on the same operation types, but with different numbers of operations---FPOs are almost perfectly correlated with energy and runtime efficiency. 
Thus, while FPOs are useful for measuring relative ordering within architecture classes, they are not adequate on their own to measure energy or even runtime efficiency. 
\end{exampleinline}

\subsection{Estimates with Partial Information Can Be Inaccurate} 
\label{sec:bad_estimates}
The current state of accounting for energy and carbon varies across fields and papers (see Section~\ref{sec:background}). Few works, if any, report all of the metrics that our framework collects.
However, it is possible to extrapolate energy and carbon impacts from some subsets of these metrics. 
This can give a very rough approximation of the energy used by an experiment in kWh (see Section~\ref{sec:background} for background). 

\begin{figure}[!htbp]
    \centering
    \includegraphics[width=.49\textwidth,valign=t]{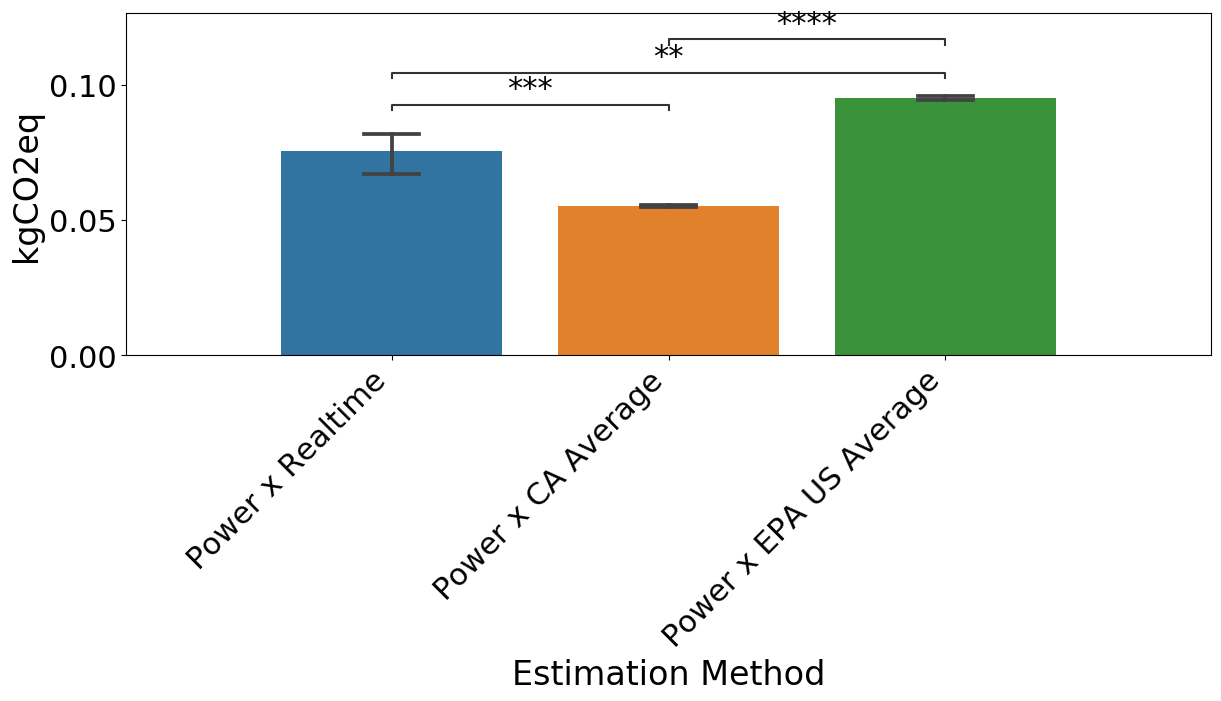}
    \includegraphics[width=.49\textwidth,valign=t]{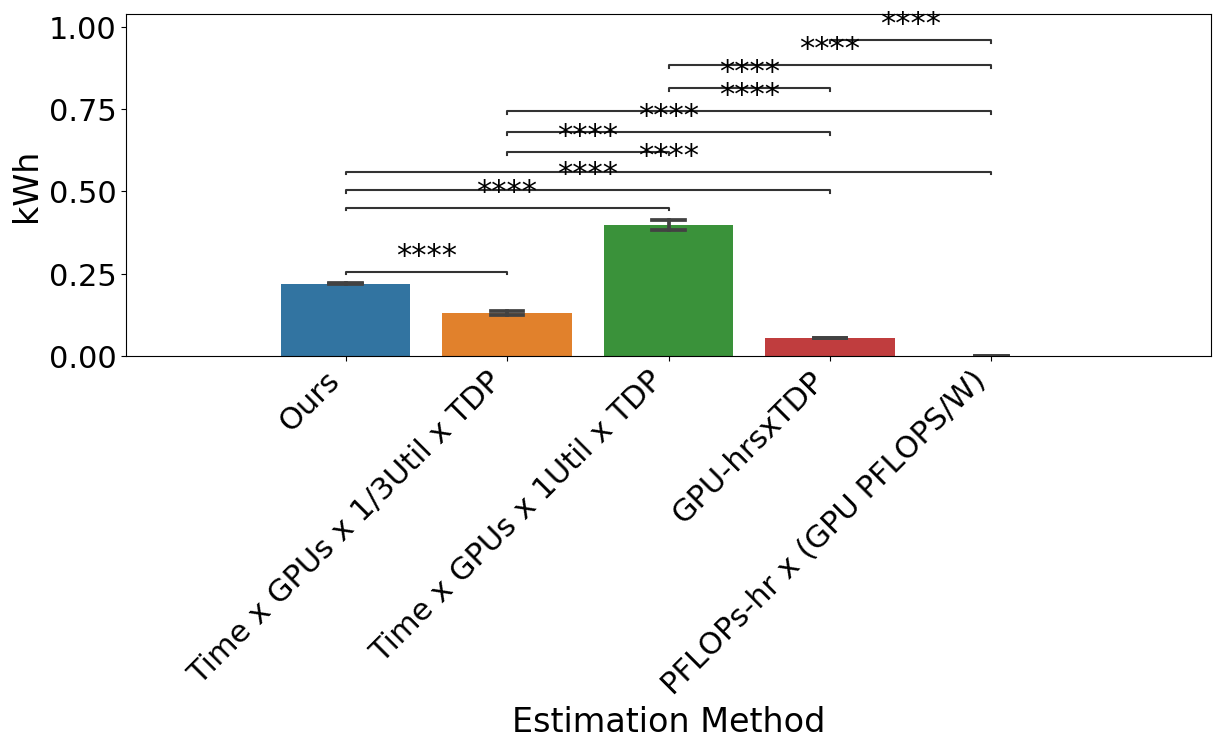}
    \caption{We compare carbon emissions (left) and kWh (right) of our Pong PPO experiment (see Appendix~\ref{app:fig2} for more details) by using different estimation methods. By only using country wide or even regional average estimates, carbon emissions may be over or under-estimated (respectively). Similarly, by using partial information to estimate energy usage (right, for more information about the estimation methods see Appendix~\ref{app:fig2}), estimates significantly differ from when collecting all data in real time (as in our method). Clearly, without detailed accounting, it is easy to over- or under-estimate carbon or energy emissions in a number of situations. Stars indicate level of significance: * p < .05, ** p < .01, *** p < .001, **** p < .0001. Annotation provided via: \href{https://github.com/webermarcolivier/statannot}{https://github.com/webermarcolivier/statannot}.}
    \label{fig:estimations}
\end{figure}

\begin{exampleinline}
We demonstrate how several such estimation methods compare against the more fine-grained accounting methods we describe in Section~\ref{sec:tool}.\footnote{We also provide a script to do the rough calculation of energy and carbon footprints based on GPU type, IP address (which is used to retrieve the location of the machine and that region's carbon intensity), experiment length, and utilization factor. \href{https://github.com/Breakend/experiment-impact-tracker/blob/master/scripts/get-rough-emissions-estimate}{https://github.com/Breakend/experiment-impact-tracker/blob/master/scripts/get-rough-emissions-estimate}}  As seen in Figure~\ref{fig:estimations}, we find significant differences from when we track all data (as through the \emph{experiment-impact-tracker} framework) to when we use partial data to extrapolate energy and carbon emissions. 
Only using GPUs and the experiment time ignores memory or CPU effects; only using the average case US region ignores regional differences. 
More details for this experiment can be found in Appendix~\ref{app:fig2}.
\end{exampleinline}

We also note that the possible estimation differences in Figure~\ref{fig:estimations} do not include possible errors from counting multiple processes at once, as described in Section~\ref{sec:per-process}.
Clearly, without detailed accounting, it is easy to severely over- or
underestimate carbon or energy emissions by extrapolating from partial information.

\section{Encouraging Efficiency and Mitigating Carbon Impacts: Immediate Mitigation Strategies}
\label{sec:mitigation}

With \emph{experiment-impact-tracker}, we hope to ease the burden of standardized reporting. We have demonstrated differences in more detailed estimation strategies from the current status quo. In this Section, we examine how accurate reporting can be used to drive immediate mitigating strategies for energy consumption and carbon emissions.

\subsection{Energy Efficiency Leaderboards} 
\label{sec:leaderboards}

A body of recent work has emphasized making more computationally efficient models~\citep{DBLP:journals/corr/abs-1901-10430,zhou2020hulk,reddi2020mlperf,lu2018low,Coleman:2019:ADT:3352020.3352024,2019arXiv191000762J}, yet another line of work has focused on the opposite: building larger models with more parameters to tackle more complex tasks~\citep{aiandcompute,sutton2019bitter}. 
We suggest leaderboards which utilize carbon emissions and energy metrics to promote an informed balance of performance and efficiency. 
DawnBench~\citep{DBLP:journals/corr/abs-1901-10430}, MLPerf~\citep{reddi2020mlperf}, and HULK~\citep{zhou2020hulk} have done this in terms of runtime and cost. \citet{ethayarajh2020utility} have recently critiqued leaderboards for only optimizing for one particular metric. By optimizing for energy and carbon emissions directly in addition to target performance metrics, baseline implementations can converge to more efficient climate-friendly settings. 
This can also help spread information about the most energy and climate-friendly combinations of hardware, software, and algorithms such that new work can be built on top of these systems instead of more energy-hungry configurations.\footnote{Something to note is that we do not compare carbon efficiency directly---instead focusing on energy specifically. Since running at different times of day and in different regions can affect carbon impacts, these may not have anything to do with the algorithm hardware-software stack and increase the number of confounds when comparing algorithms. While hardware is also immutable to some extent, there may still be information to be gained by finding combinations of efficient low-level optimizations for specific hardware. Hardware can also be held relatively constant by using the same machine for all experimental runs. If comparisons using carbon units are desired, a fixed carbon intensity factor should likely be chosen for approximate comparisons in a given region (rather than using live carbon intensity metrics). See, also, Appendix~\ref{app:comparisons}.}

\begin{figure}
    \centering
    \includegraphics[width=.49\textwidth]{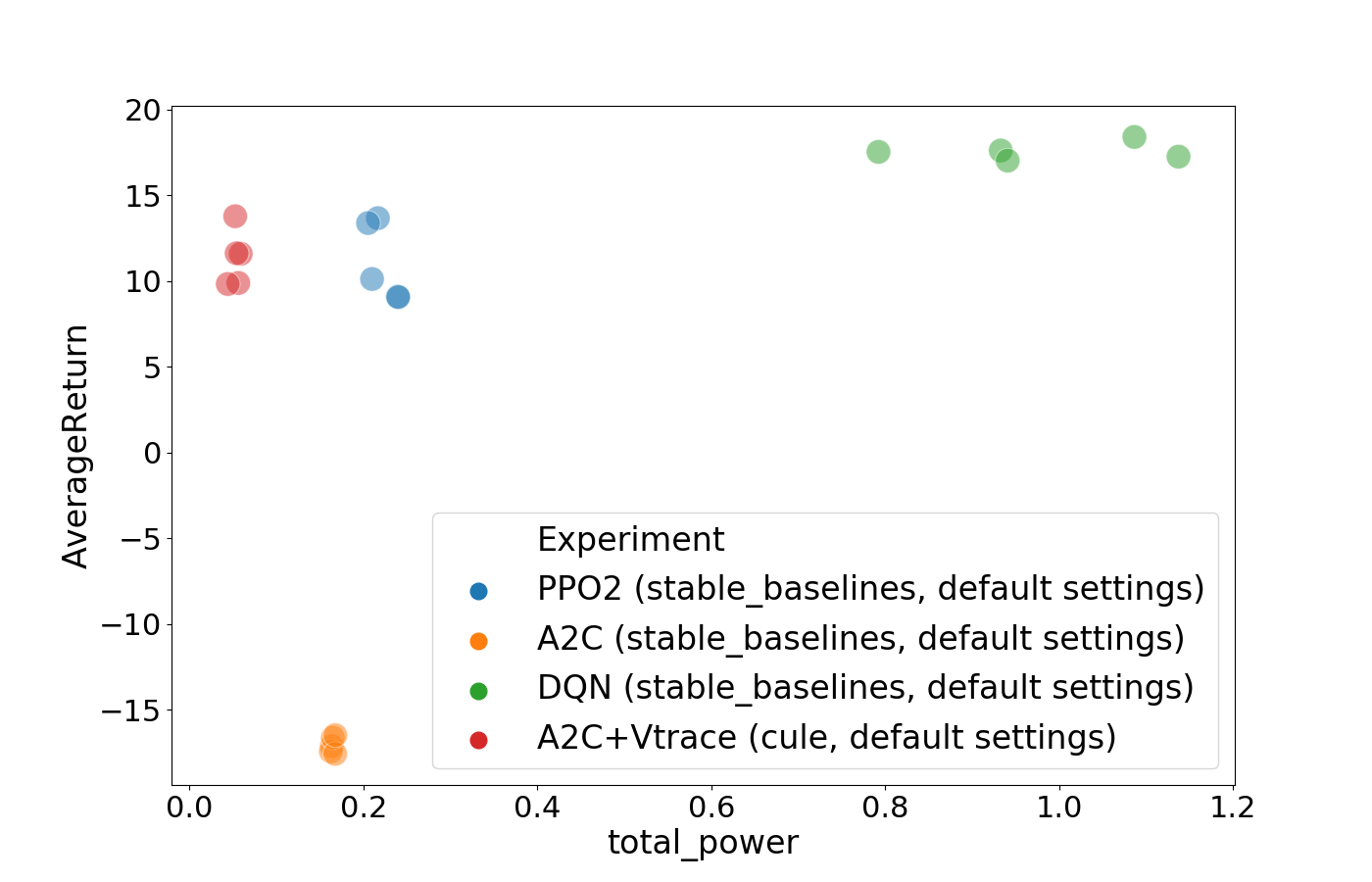}
    \includegraphics[width=.49\textwidth]{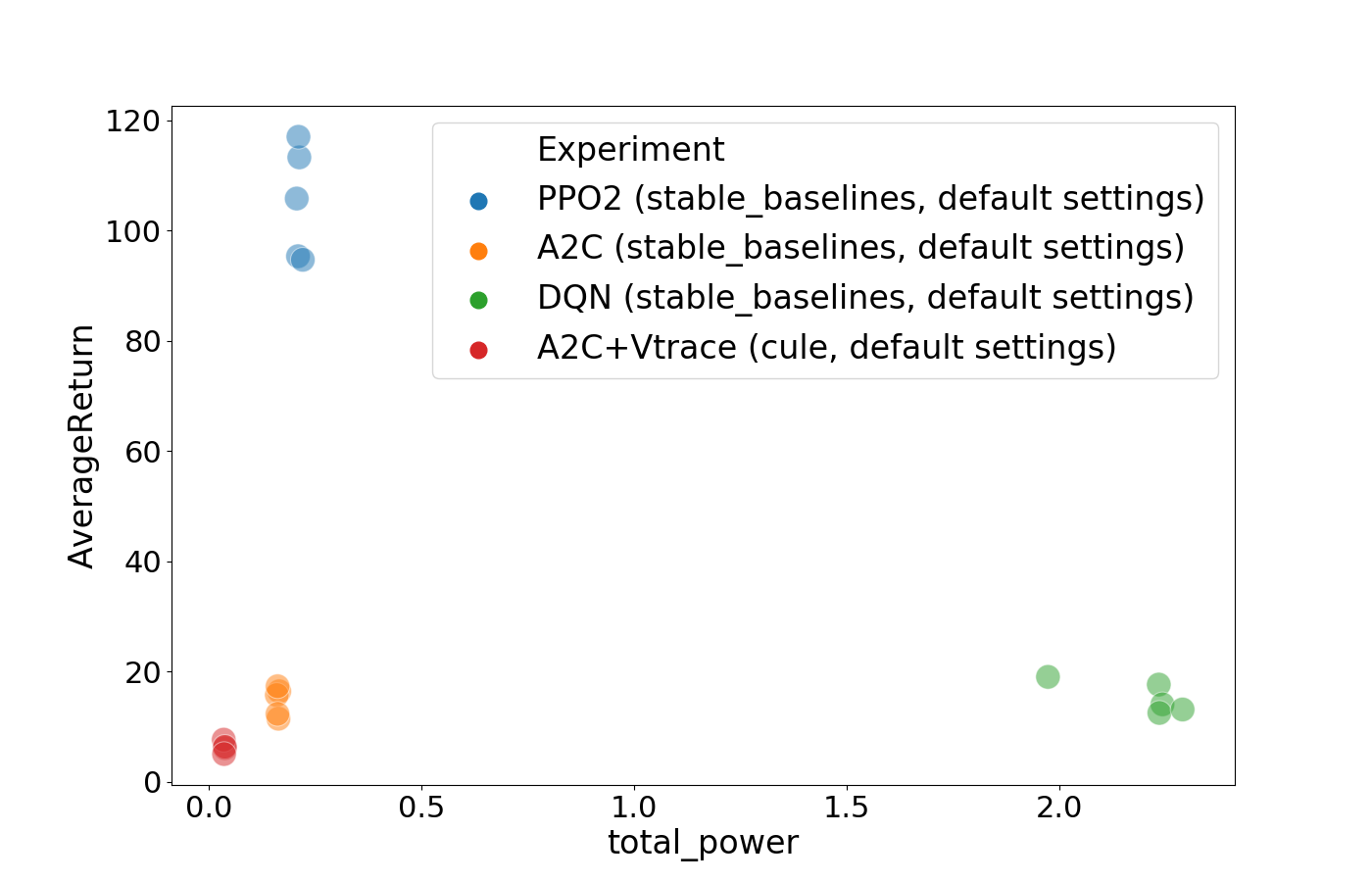}
    \caption{We evaluate A2C, PPO, DQN, and A2C+VTraces on PongNoFrameskip-v4 (left) and BreakoutNoFrameskip-v4 (right), two common evaluation environments included in OpenAI Gym. 
    We train for only 5M timesteps, less than prior work, to encourage energy efficiency and evaluate for 25 episodes every 250k timesteps. We show the Average Return across all evaluations throughout training (giving some measure of both ability and speed of convergence of an algorithm) as compared to the total energy in kWh. 
    Weighted rankings of Average Return per kWh place A2C+Vtrace first on Pong and PPO first on Breakout. Using PPO versus DQN can yield significant energy savings, while retaining performance on both environments (in the 5M samples regime). 
    See Appendix~\ref{app:rl} for more details and results in terms of asymptotic performance.}
    \label{fig:rl_average}
\end{figure}

\subsubsection{A Deep RL Energy Leaderboard}
\label{sec:rl}

To demonstrate how energy leaderboards can be used to disseminate information on energy efficiency, we create a Deep RL Energy Leaderboard.\footnote{\url{https://breakend.github.io/RL-Energy-Leaderboard/reinforcement_learning_energy_leaderboard/index.html}} The website is generated using the same tool for creating HTML appendices described in Section~\ref{sec:tool}. All information (except for algorithm performance on tasks) comes from the \emph{experiment-impact-tracker} framework. We populate the leaderboard for two common RL benchmarking environments, PongNoFrameskip-v4 and BreakNoFrameskip-v4~\citep{bellemare2013arcade,1606.01540,mnih2013playing}, and four baseline algorithms, PPO \citep{schulman2017proximal}, A2C~\citep{mnih2016asynchronous}, A2C with V-Traces~\citep{espeholt2018impala,dalton2019gpuaccelerated}, and DQN~\citep{mnih2013playing}. The experimental details and results can also be found in Figure~\ref{fig:rl_average}. We find that no algorithm is the energy efficiency winner across both environments, though the PPO implementation provided by \citet{stable-baselines} attains balance between efficiency and performance when using default settings across algorithms.
\begin{exampleinline}
To see how such a leaderboard might help save energy, consider a Deep RL class of 235 students.\footnote{See for example, \href{https://explorecourses.stanford.edu/search?q=deep+reinforcement+learning&view=catalog&page=0&filter-coursestatus-Active=on&collapse=&academicYear=20182019}{Stanford's CS 234}.} For a homework assignment, each student must run an algorithm 5 times on Pong. The class would save 888 kWh of energy by using PPO versus DQN, while achieving similar performance.\footnote{These rankings may change with different code-bases and hyperparameters.} This is roughly the same amount needed to power a US home for one month.\footnote{\href{https://web.archive.org/web/20191214183939/https://www.eia.gov/tools/faqs/faq.php?id=97&t=3}{https://www.eia.gov/tools/faqs/faq.php?id=97\&t=3}} 
\end{exampleinline}

We, thus, encourage the community to submit more data to the leaderboard to find even more energy efficient algorithms and configurations. 

\subsection{Running In Carbon-Friendly Regions} 
\label{sec:regional_carbon}
\begin{figure}
    \centering
    \includegraphics[width=\textwidth]{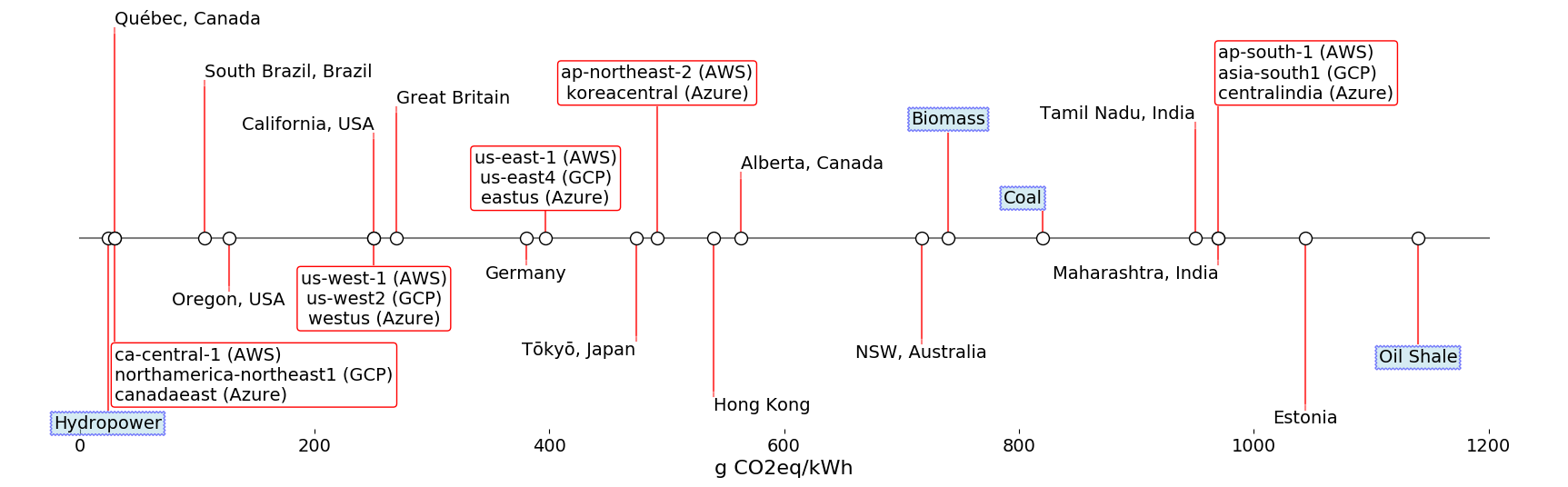}
    \caption{Carbon Intensity (g\coeq/kWh) of selected energy grid regions is shown from least carbon emissions (left) to most carbon emissions (right). Red/unshaded boxes indicate carbon intensities of cloud provider regions. Blue/shaded boxes indicate carbon intensities of various generation methods. Oil shale is the most carbon emitting method of energy production in the Figure. Estonia is powered mainly by oil shale and thus is close to it in carbon intensity. Similarly, Québec is mostly powered by hydroelectric methods and is close to it in carbon intensity. Cloud provider carbon intensities are based on the regional energy grid in which they are located. Thus, us-west-1, located in California, has the same carbon intensity as the state. See \url{https://github.com/Breakend/experiment-impact-tracker/} for data sources of regional information. Energy source information from \citet{iiasa11109,iea}. }
    \label{fig:regional_differences}
\end{figure}

We noted in Section~\ref{sec:tool} that it is important to assess which energy grid experiments are run on due to the large differences in carbon emissions between energy grids. Figure~\ref{fig:regional_differences} shows \coeq intensities for an assortment of locations, cloud-provider regions, and energy production methods. We note that an immediate drop in carbon emission can be made by moving all training jobs to carbon-efficient energy grids. In particular, Quebec is the cleanest available cloud region to our knowledge. Running a job in Quebec would result in carbon emission 30x lower than running a job in Estonia (based on 2017 averages). 

\begin{figure}
    \centering
    \includegraphics[scale=.3,valign=t]{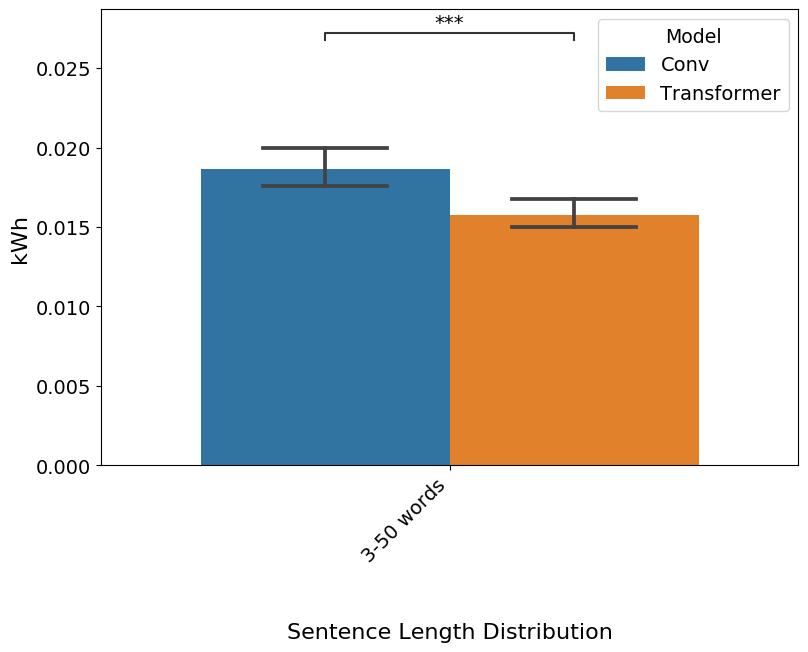}
    \includegraphics[scale=.3,valign=t]{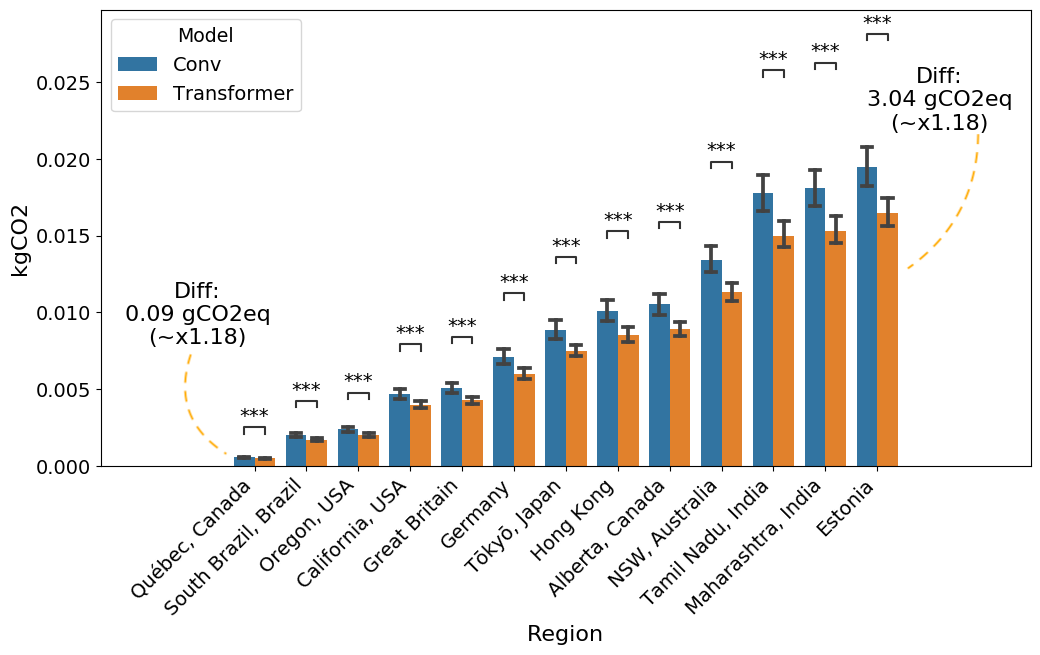}
    \caption{We use pre-trained En-Fr translation models downloaded from PyTorch Hub: a convolutional network~\citep{gehring2017convolutional} and transformer~\citep{ott2018scaling}. We generate 1000 random sequences either between 3-50 words in length using the essential\_generators Python package: \url{https://pypi.org/project/essential-generators/}. We repeat with 20 random seeds. Randomly generated sentences are likely to be difficult to translate, but this difficulty should not be biased in favor of either algorithm. [Left] We show the true difference in energy consumption. [Right] We show estimated kg\coeq  released if the experiment had been conducted in a number of increasingly carbon-intensive energy grids. Differences remain significant throughout, but the absolute difference increases as more carbon-intensive regions are assumed.}
    \label{fig:transformer}
\end{figure}

\begin{exampleinline}
To demonstrate this in practice, we run inference on two machine translation models 1000 times and measure energy usage. We extrapolate the amount of emissions and the difference between the two algorithms if run in different energy grids, seen in Figure~\ref{fig:transformer}. The absolute difference in emissions between the two models is fairly small (though significant) if run in Quebec (.09 g \coeq), yet the gap increases as one runs the jobs in less carbon-friendly regions (at 3.04 g \coeq \, in Estonia).
\end{exampleinline} 

We provide a script with our framework to show all cloud provider region with emission statistics to make this decision-making process easier.\footnote{See: \href{https://github.com/Breakend/experiment-impact-tracker/blob/master/scripts/get-region-emissions-info}{get-region-emissions-info script} and \href{https://github.com/Breakend/experiment-impact-tracker/blob/master/scripts/lookup-cloud-region-info}{lookup-cloud-region-info script}.} We note that \citet{lacoste2019quantifying} provide a website using partial information estimation to extrapolate carbon emissions based on cloud provider region, GPU type, and experiment length in hours. Their tool may also be used for estimating carbon emissions in cloud-based experiments ahead of time. We've also provided a non-exhaustive list of low emissions energy grids that contain cloud regions in Table~\ref{tab:cloudregions}.

For companies that train and deploy large models often, shifting these resources is especially important.
ML training is not usually latency bound: companies can run training in cloud regions geographically far away since training models usually does not require round trip communication requirements.
Contrary to some opinions,\footnote{\href{https://web.archive.org/web/20191218223658/https://www.theguardian.com/technology/2019/sep/17/tech-climate-change-luddites-data}{https://www.theguardian.com/technology/2019/sep/17/tech-climate-change-luddites-data}} there is not a necessary need to eliminate computation-heavy models entirely, as shifting training resources to low carbon regions will immediately reduce carbon emissions with little impact to production systems.
For companies seeking to hit climate change policy targets, promotion of carbon neutral regions and shifting of all machine learning systems to those regions would accelerate reaching targets significantly and reduce the amount of offset purchasing required to meet goals (thus saving resources).\footnote{See, for example, Amazon's goal: \href{https://web.archive.org/web/20191213130014/https://press.aboutamazon.com/news-releases/news-release-details/amazon-co-founds-climate-pledge-setting-goal-meet-paris/}{https://press.aboutamazon.com/news-releases/news-release-details/amazon-co-founds-climate-pledge-setting-goal-meet-paris}} It is worth noting that some companies like Google already purchase offsets~\citep{google_rec1}, so it may be unclear why shifting resources is necessary.
We provide an extended discussion on this in Appendix~\ref{app:carbon_discussion}. As a matter of total emissions reductions, running compute in carbon-friendly regions prevents emissions now, while offsets may not come into effect for several years. Moreover, continuing offset purchasing at current levels, while shifting resources to green regions would result in a net-negative carbon footprint.

\begin{table}
    \centering
    \begin{tabular}{p{5cm}|p{5cm}|p{4cm}}
    \toprule
        Power Grid & Cloud Regions & Carbon Intensity \\
        &&(g \coeq / kWh)\\
        \midrule
         Quebec, Canada & ca-central-1 (AWS),& $\sim$ 30\\ 
         & canadaeast (Azure), & \\
         &northamerica-northeast1 (GCP)&\\ 
         \midrule
         West Norway & norwaywest (Azure) & $\sim$ 35\\
                  \midrule
         Ontario, Canada & canadacentral (Azure) & $\sim$ 45\\

        \midrule
         France & eu-west-3 (AWS), francesouth (Azure), francecentral (Azure) & $\sim$ 56 \\

         \midrule
         Brazil (Central) & brazilsouth (Azure) & $\sim$ 106\\
                  \midrule
         Oregon, USA & us-west1 (GCP),  & $\sim$ 127\\
         & us-west-2 (AWS)&\\
         & westus2 (Azure) &\\
         \bottomrule
    \end{tabular}
    \caption{A non-exhaustive list of cloud regions in low carbon intensity energy grids ($<150$ g\coeq / kWh). All estimates pulled as yearly averages from \url{https://www.electricitymap.org/map}, except for Quebec which utilizes methodology from \url{https://piorkowski.ca/rev/2017/06/canadian-electricity-co2-intensities/} and Oregon which uses data from \url{https://www.eia.gov/electricity/state/oregon/}.}
    \label{tab:cloudregions}
\end{table}

\section{Discussion: Systemic Changes}

We demonstrated several use cases for accounting which can drive immediate mitigation strategies. However, the question remains: how can we encourage systemic changes which lead to energy and carbon efficiency in ML systems?

\subsection{Green Defaults for Common Platforms and Tools} 
\label{sec:ease-of-use}

Energy leaderboards help provide information on energy efficient configurations for the whole stack. However, to truly spread energy efficient configurations, underlying frameworks should by default use the most energy-efficient settings possible. This has been shown to be an effective way to drive pro-environmental behavior~\citep{pichert2008green}. For example, Nvidia apex provides easy mixed-precision computing as an add-on which yields efficiency gains.\footnote{\url{https://github.com/NVIDIA/apex}} However, it requires knowing this and using it. \citet{merity2019single} also discusses the current difficulties in using highly efficient components. Making such resources supported as defaults in frequently used frameworks, like PyTorch, would immediately improve the efficiency of all downstream projects. We encourage maintainers of large projects to prioritize and support such changes.

\subsection{How Much Is Your Performance Gain Worth? Balancing Gains With Cost} 
\label{sec:cost-benefit}

While training jobs can easily be shifted to run in clean regions, there are often restrictions for inference-time use of machine learning models which prevent such a move. Many companies are deploying large machine learning models powered by GPUs for everyday services.\footnote{See for example, search which now uses transformer networks at both Microsoft and Google. \href{https://web.archive.org/web/20191226085336/https://blog.google/products/search/search-language-understanding-bert/}{https://www.blog.google/products/search/search-language-understanding-bert/} and \href{https://web.archive.org/web/20191224113641/https://azure.microsoft.com/en-us/blog/microsoft-makes-it-easier-to-build-popular-language-representation-model-bert-at-large-scale/}{https://azure.microsoft.com/en-us/blog/microsoft-makes-it-easier-to-build-popular-language-representation-model-bert-at-large-scale/}} 

\begin{exampleinline}
Production machine translation services, can process 100B words per day~\citep{turovsky2016ten}: roughly 4.2 million times our experiment in Figure~\ref{fig:transformer}.
If all translation traffic were in Estonia, 12,768 kg\coeq~(the carbon sequestered by 16.7 acres of forest in one year~\citep{us2008greenhouse}) would be saved per day by using the more efficient model, yet if all traffic were in Québec, 378 kg\coeq~would be saved (the carbon sequestered by .5 acres of forest in one year~\citep{us2008greenhouse}).
Considering the amounts of required compute, small differences in efficiency can scale to large emissions and energy impacts.
\end{exampleinline}

These services are latency-bound at inference time and thus cannot mitigate carbon emissions by shifting to different regions. Instead, deploying energy-efficient models not only reduces carbon emissions but also benefits the companies by bringing the energy costs down. We encourage companies to consider weighing energy costs (both social and monetary) with the performance gains of a new model before deploying it. In the case of our translation experiment in Figure~\ref{fig:transformer}, the pre-trained convolutional model we use is significantly more energy hungry across runs than the transformer model we use. When deploying a new energy-hungry translation model, we ask companies to consider: is the BLEU score improvement really worth the energy cost of deploying it?  Are there ways to route to different models to balance this trade-off? For example, suppose an energy-hungry model only improves performance in some subset of the data. Routing to this model only in that subset would maximize performance while minimizing energy footprint.\footnote{Efficient routing of traffic to regions has been considered before by \citet{nguyen2012environmental} and \citet{berral2010towards}. It may be worth considering efficient routing of traffic to particular models as well.}. 

We note that considering such trade-offs is of increased importance for models aiming to reduce carbon emissions as described by~\citet{rolnick2019tackling}. 
Deploying a large deep learning model for, say, improving the energy efficiency of a building, is not worth it if the energy costs of the model outweigh the gains.
We also leave an open question to economists to help assess the welfare benefits of gains on a particular machine learning metric (e.g., how much is BLEU score worth in a translation service). This would allow the social welfare of the metric to be balanced against the social cost of carbon~\citep{countrylevelsocialcostofcarbon} for deployment decisions. 

Similarly, it is important to consider other types of cost-benefit analyses. Perhaps the carbon impacts of a long (energy-intensive) training time for a large model is worth it if it reduces the lifetime carbon footprint in production (for example, if the model doesn't require expensive fine-tuning procedures in the future).
Understanding the tradeoff between the lifetime deployment costs and training costs is important before moving on to extended training runs. As such, we also encourage reporting of both estimated training and deployment energy costs so future adopters have a more comprehensive picture when deciding which model to use.

Central to all of these cost-benefit analyses are accurate accounting. Our tool provides one step in consistent and accurate accounting for such purposes.

\subsection{Efficient Testing Environments}
\label{sec:efficiency-test-env}
In Section~\ref{sec:ease-of-use} we discuss the adoption of green default configurations and Section~\ref{sec:cost-benefit} discusses cost-benefit analyses for deployments. Another consideration particular to research---especially RL---is the selection of the most efficient testing environments which assess the mechanism under test. For example, if an RL algorithm solves a particularly complex task in an interesting way, like solving a maze environment, is there a way to demonstrate the same phenomenon in a more efficient environment? Several works have developed efficient versions of RL environments which reduce run-times significantly. In particular, \citet{dalton2019gpuaccelerated} improve the efficiency of Atari experiments by keeping resources on the GPU (and thus avoiding energy and time overheads from moving memory back and forth). \citet{gym_minigrid} develop a lightweight Grid World environment with efficient runtimes for low-overhead experiments. An important cost-benefit question for researchers is whether the same point can be proven in a more efficient setting. 

\subsection{Reproducibility} 
\label{sec:reproducibility}
A key aspect to our work is helping to promote reproducibility by aiding in consistent reporting of experimental details. We encourage all researchers to release code and models (when it is socially and ethically responsible to do so), to prevent further carbon emissions. Replicating results is an important, if not required, part of research. If replication resources are not available, then more energy and emissions must be spent to replicate results---in the case of extremely large models, the social cost of carbon may be equivalently large. 
Thus, we ask researchers to also consider energy and environmental impacts from replication efforts, when weighing model and code release. 
We note that there may very well be cases where safety makes this trade-off lean in the direction of withholding resources, but this is likely rare in most current research. For production machine learning systems, we encourage developers to release models and codebases internally within a company. This may encourage re-use rather than spending energy resources developing similar products.

\subsection{Standardized Reporting} 
\label{sec:standard}

We suggest that all papers include standardized reporting of energy and carbon emissions. We also suggest adding a Carbon Impact Statement at the end of papers (just like ours below) which estimates the carbon emissions of the paper. This can be reported in a dollar amount via the country-specific social cost of carbon~\citep{countrylevelsocialcostofcarbon}. We provide a script\footnote{\url{https://github.com/Breakend/experiment-impact-tracker/blob/master/scripts/generate-carbon-impact-statement}} to parse logs from the \emph{experiment-impact-tracker} framework and generate such a statement automatically. We suggest this to spread awareness and bring such considerations to the forefront. We encourage this statement to include \emph{all} emissions from experimentation to build a more realistic picture of total resources spent. 

We also emphasize that research, even when compute intensive, is immensely important for progress. It is unknown what sequence of papers may inspire a breakthrough~\citep{stanley2015greatness} which would reduce emissions by more than any suggestion here. While emissions should be minimized when possible, we suggest that impact statements be only used for awareness. This is especially true since access to clean energy grids or hardware may be limited for some in the community.

We also suggest that, when developing features which visualize compute intensity for cloud or internal workloads, developers consider providing built-in tools to visualize energy usage and carbon emissions. For example, the Colab Research Environment shows RAM and Disk capacity,\footnote{\href{https://colab.research.google.com/}{https://colab.research.google.com/}} but could also show and provide access to these other metrics more easily. Providing similar informational labels~\citep{byerly2018nudging} within internal tooling could mitigate some energy and carbon impacts within companies.

\subsection{Badging}
\label{sec:badging}
Informational labeling has had a long history of being used in public policy~\citep{banerjee2003eco}. In the USA, the ``Energy Star'' label has been used to guide customers to eco-friendly products. More recently, ``badges'' rewarded  by the \emph{Psychological Science} journal were shown to be effective, with a jump from 3\% of articles reporting open data to 39\% one year later. ACM has introduced similar reproducibility badges.\footnote{\href{https://web.archive.org/web/20191209171518/https://www.acm.org/publications/policies/artifact-review-badging}{https://www.acm.org/publications/policies/artifact-review-badging}} With consistent reporting of carbon and energy metrics, climate friendly research badges can be introduced by conferences to recognize any paper that demonstrates a significant effort to mitigate its impacts. For example, a compute intensive paper, when showing evidence of explicitly running resources in a clean region can be rewarded with such a badge. Another example badge can be awarded to papers that create energy-friendly algorithms with similar performance as the state-of-the-art\footnote{See, for example, \citet{clark2020electra} which creates a more efficient version of text encoder pre-training.}. The goal of these badges is to draw further attention to efficient versions of state-of-the-art systems and to encourage mitigation efforts while, again, not punishing compute-intensive experiments. Of course this may not apply to conferences such as SysML which often focus on making models more efficient, but rather as a motivational tool for other venues where efficiency may not be in focus.

\subsection{Limitations and Opportunities for Extensions} 
\label{sec:drivers}
The \emph{experiment-impact-tracker} framework abstracts away many of the previously mentioned difficulties in estimating carbon and energy impacts: it handles routing to appropriate tools for collecting information, aggregates information across tools to handle carbon calculations, finds carbon intensity information automatically, and corrects for multiple processes on one machine. Yet, a few other challenges may be hidden by using the framework which remain difficult to circumvent.

As \citet{Khan:2018:RAE:3199681.3177754} discuss, and we encounter ourselves, poor driver support makes tracking energy difficult. Not every chipset supports RAPL, nor does every Linux kernel. Intel also does not provide first party supported python libraries for access to measurements.
\emph{nvidia-smi} per-process measurements in docker containers are not supported.\footnote{\href{https://web.archive.org/web/20191226201234/https://github.com/NVIDIA/nvidia-docker/issues/179/\%23issuecomment-242150861}{https://github.com/NVIDIA/nvidia-docker/issues/179\#issuecomment-242150861}} 
A body of work has also looked at improving estimates of energy usage from RAPL by fitting a regression model to real energy usage patterns~\citep{DBLP:journals/corr/abs-1709-06076,kavanagh2019rapid,ghosh2013statistical,song2013unified}. 
The Slurm workload manager provides an energy accounting plugin,\footnote{\href{https://web.archive.org/web/20160806015613/http://slurm.schedmd.com/acct_gather_energy_plugins.html}{https://slurm.schedmd.com/acct\_gather\_energy\_plugins.html}} but requires administrator access to add. For those without access to Slurm, Intel's RAPL supports access to measurements through three mechanisms, but only one of these (the powercap interface only available on some systems) does not require root access (see more discussion by \citet{Khan:2018:RAE:3199681.3177754}). To promote widespread reporting, we avoid any tool which requires administrative access or would not be accessible on most Linux systems.
Providing better supported tools for user-level access to power metrics would make it possible to more robustly measure energy usage. Aggregating metrics and handling the intricacies of these downstream tools requires time and knowledge. We try to abstract as much of these challenges away in the \emph{experiment-impact-tracker}, though some driver-related issues require driver developer support. 
However, these issues make it difficult to support every on-premises or cloud machine. As such, we currently only support instances which have Intel RAPL or PowerGadget capabilities for Mac OS and Linux. 

We also note that carbon intensities for machines in cloud data centers may not reflect the regional carbon intensities. Some providers buy clean energy directly for some data centers, changing the realtime energy mix for that particular data center. We were unable to find any information regarding realtime energy mixes in such cases and thus could not account for these scenarios. If providers exposed realtime APIs for such information this would help in generating more accurate estimates. Moreover, customized hardware in cloud provider regions does not always provide energy accounting mechanisms or interfaces. If cloud providers supported libraries for custom hardware, this could be used for more detailed accounting in a wider range of cloud-based compute scenarios.

We further discuss other sources of error and issues arising from these difficulties in Appendix~\ref{sec:sourcesoferror}.

\section{Concluding Remarks and Recommendations}

We have shown how the \emph{experiment-impact-tracker} and associated tools can help ease the burden of consistent accounting and reporting of energy, compute, and carbon metrics; we encourage contribution to help expand the framework. We hope the Deep RL Energy Leaderboard helps spread information on energy efficient algorithms and encourages research in efficiency. 
While we focus on compute impacts of machine learning production and research, a plethora of other work considers costs of transportation for conferences~\citep{holden2017academic,spinellis2013carbon,bossdorf2010climate} and compute hardware manufacturing~\citep{venkatesan2015comparative}. We encourage researchers and companies to consider these other sources of carbon impacts as well. Finally, we recap several points that we have highlighted in mitigating emissions and supporting consistent accountability.
\newpage

\noindent \emph{What can machine learning researchers do?}

\begin{itemize}
    \item Run cloud jobs in low carbon regions only (see Section~\ref{sec:regional_carbon}).
    \item Report metrics as we do here, make energy-efficient configurations more accessible by reporting these results (see Section~\ref{sec:standard}).
    \item Work on energy-efficient systems, create energy leaderboards (see Section~\ref{sec:mitigation}).
    \item Release code and models whenever safe to do so (see Section~\ref{sec:reproducibility}).
    \item Integrate energy efficient configurations as defaults in baseline implementations (see Section~\ref{sec:ease-of-use}).
    \item Encourage climate-friendly initiatives at conferences (see Sections~\ref{sec:badging} and \ref{sec:standard}).
\end{itemize}

\noindent \emph{What can industry machine learning developers and framework maintainers do?}

\begin{itemize}
    \item Move training jobs to low carbon regions immediately. Make default launch configurations and documentation point to low carbon regions (see Section~\ref{sec:regional_carbon}).
    \item Provide more robust tooling for energy tracking and carbon intensities (see Section~\ref{sec:drivers}).
    \item Integrate energy efficient operations as default in frameworks (see Section~\ref{sec:ease-of-use}).
    \item Release code and models (even just internally in the case of production systems) whenever safe to do so (see Section~\ref{sec:reproducibility}). 
    \item Consider energy-based costs versus benefits of deploying new models (see Section~\ref{sec:cost-benefit}).
    \item Report model-related energy metrics (see Section~\ref{sec:standard}).
\end{itemize}

We hope that regardless of which tool is used to account for carbon and energy emissions, the insights we provide here will help promote responsible machine learning research and practices. 

\section*{Carbon Impact Statement}
This work contributed 8.021 kg of $\text{CO}_{2eq}$ to the atmosphere and used 24.344 kWh of electricity, having a USA-specific social cost of carbon of \$0.38 (\$0.00, \$0.95). Carbon accounting information located at: \url{https://breakend.github.io/ClimateChangeFromMachineLearningResearch/measuring_and_mitigating_energy_and_carbon_footprints_in_machine_learning/} and \url{https://breakend.github.io/RL-Energy-Leaderboard/reinforcement_learning_energy_leaderboard/index.html}. The social cost of carbon uses models from \citet{countrylevelsocialcostofcarbon}. This statement and carbon emissions information was generated using \emph{experiment-impact-tracker} described in this paper.

\bibliography{references}  %

\appendix

\section{Conference Travel}
\label{sec:conference_travel}
Prior work has also examined conference travel for various fields as a major source of impact~\cite{spinellis2013carbon,astudillo2018estimating,psychimpact}. For example, \citet{spinellis2013carbon} found that the \coeq emissions from travel per conference participant was about 801 kg \coeq, \citet{astudillo2018estimating} estimated around 883 kg \coeq~emissions per participant, and \citet{psychimpact} estimate around 910 kg of \coeq~emissions per participant. Interestingly, these separate papers all align around the same carbon emissions numbers per conference participant. Using this and ML conference participant statistics we can gain some (very) rough insight into the carbon emissions caused by conference travel (not including food purchases, accommodations, and travel within the conference city).

Conference participation has grown particularly popular in ML research, attracting participants from industry and academia. In 2018 the Neural Information Processing Systems (NeurIPS) conference sold out registrations in 12 minutes~\citep{shead_2018}. In 2019, according to the AI Index Report 2019~\citep{shoham2018ai}, conferences had the following attendance: CVPR (9,227); IJCAI (3,015); AAAI (3,227); NeurIPS (13,500); IROS (3,509); ICML (6,481); ICLR (2,720); AAMAS (701); ICAPS (283); UAI (334). The larger conferences also showed continued growth: NeurIPS showed a year-over-year growth 41\% from 2018 to 2019. Given only these conferences and their attendances in 2019, the lower 801kg \coeq~average emissions estimate per participant~\citep{spinellis2013carbon}, this adds up to roughly 34,440,597 kg \coeq~emitted in 2019 from ML-related conferences (not considering co-location and many other factors).

\section{NeurIPS Sampling on Metric Reporting}
\label{app:neurips}

We randomly sampled 100 NeurIPS papers from the 2019 proceedings, of these papers we found 1 measured energy in some way, 45 measured runtime in some way, 46 provided the hardware used, 17 provided some measure of computational complexity (e.g., compute-time, FPOs, parameters), and 0 provided carbon metrics. We sampled from the NeurIPS proceedings page: \url{https://papers.nips.cc/book/advances-in-neural-information-processing-systems-32-2019}. We first automatically check for key words (below) related to energy, compute, and carbon. We then examined the context of the word to classify it as relating to hardware details (e.g., Nvidia Titan X GPU), computational efficiency (e.g., FPOs, MAdds, GPU-hours), runtime (e.g., the experiment ran for 8 hours), energy (e.g., a plot of performance over Joules or Watts), or carbon (e.g., we estimate 10 kg \coeq were emitted). We also manually validate papers for similar metrics that didn't appear in the keyword search. If a paper did not contain experiments we removed it and randomly redrew a new paper. In many cases, metrics are only provided for some subset of experiments (or for particular ablation experiments). We nonetheless count these as reporting the metric. Where a neural network diagram or architecture description was provided, we did not consider this to be reporting a compute metric.

compute\_terms = ["flop", "fpo", "pflop", "tflops", "tflop", "parameters", "params", "pflops", "flops", "fpos", "gpu-hours", "cpu-hours", "cpu-time", "gpu-time", "multiply-add", "madd"]

hardware\_terms = ["nvidia", "intel", "amd", "radeon", "gtx", "titan", "v100", "tpu", "ryzen", "cpu", "gpu"]

time\_terms = ["seconds", "second", "hour", "hours", "day", "days", "time", "experiment length", "run-time", "runtime"]

energy\_terms = ["watt", "kWh", "joule", "joules", "wh", "kwhs", "watts", "rapl", "energy", "power"]

carbon\_terms = ["co2", "carbon", "emissions"]

\section{Carbon Discussion}
\label{app:carbon_discussion}

\begin{center}
\emph{But cloud providers claim 100\% carbon neutrality in my region, why do I need to shift my resources?} 
\end{center}

While we estimate energy mixes based on regional grids, cloud providers sometimes aim for carbon \emph{neutrality} through a mixture of mechanisms which may change the energy mix being provided to a data center in an otherwise carbon intensive energy grid or otherwise offset unclean energy usage.
Data centers draw energy from the local energy grids and as a result the mix of energy they consume largely depends on the composition of the power running in the grids. 
If the local energy grids are powered by a mix of fuel and renewable energy, a data center will inevitably consume fuel energy as well. 

Due to the fact that the consumers do not know the origin of the physical electricity from the utility grid, it is difficult to assign ownership of the renewable energy consumption. The Environmental Protection Agency (EPA) uses renewable energy certificates (RECs) to track the generation and consumption of renewable energy: one REC is issued when one megawatt-hour (MWh) of electricity is generated from a renewable source and delivered to the energy grid.\footnote{\url{https://www.epa.gov/greenpower/renewable-energy-certificates-recs}} 
Consumers can then purchase RECs from a renewable energy provider and apply them to their electricity usage. This means consumers can claim they run on renewable energy by purchasing RECs from providers that doesn't actually power the energy grids that they draw electricity from. 
Although this means that the consumers' realtime carbon footprints will still be decided by the composition of renewable and fuel energy in their local energy grids, more renewable energy can flow onto the grid by purchasing the RECs and future development of renewable sources is supported.
Google, to offset its carbon emissions, uses RECs and power purchase agreements (PPAs) with renewable energy providers to ensure that more renewable energy powers the same electricity grids that its data centers are in.\footnote{We note that this process is likely similar for most cloud providers, but Google is the most open with their methodology, so we are able to gain more insight from the materials they publish. Information described here is mainly put together from \citet{google_rec1} and \citet{google_rec2}.} 
Google then sells the renewable energy as it becomes available back to the electricity grids and strips away the RECs. 
Over one year, Google applies equal amounts of RECs to its data centers' total energy consumption. 
This method helps green energy provider development by creating a long term demand. However, PPAs provide RECs for \emph{future renewables}, not only current energy on the grid which may remain unchanged. As it states: ``While the renewable facility output is not being used directly to power a Google data center, the PPA arrangement assures that additional renewable generation sufficient to power the data center came on line in the area.''\footnote{\url{https://static.googleusercontent.com/media/www.google.com/en/us/green/pdfs/renewable-energy.pdf}}

We can see that even if a cloud provider's data centers are carbon neutral, the actual \coeq emissions can vary largely and depends on the region and even time of the day (solar energy cannot be generated at night). Since carbon emissions have some long-term or irreversible impacts~\citep{solomon2009irreversible}, avoiding carbon emissions now can help down the line---a reason why discount rates are used in calculating impacts~\citep{weisbach2008climate}.
We suggest that cloud providers release tools for understanding the carbon intensity for each data center region regardless of offset purchasing. While the purchases of PPAs and RECs are valuable for driving towards renewable energy in otherwise dirty regions, for machine learning model training, where the resources can be moved, we believe shifting resources to low intensity regions is more beneficial to long term carbon impacts.
Other cloud-based jobs where latency requirements prevent shifting resources will remain to drive PPA/REC purchasing, and consequently renewable energy demand.

\section{ImageNet Experiments}
\label{app:imagenet}

We load pre-trained models available through PyTorch Hub (see \url{https://pytorch.org/hub})---namely AlexNet~\citep{krizhevsky2012imagenet}, DenseNet~\citep{huang2017densely}, GoogLeNet~\citep{43022}, HardNet~\citep{chao2019hardnet}, MobileNetv2~\citep{sandler2018mobilenetv2}, ShuffleNet~\citep{zhang2018shufflenet}, SqueezeNet~\citep{iandola2016squeezenet}, VGG~\citep{simonyan2014very}, and Wide ResNets~\citep{zagoruyko2016wide}. We run 50,000 rounds of inference on a single image through pre-trained image classification models and run similar analysis to \citet{canziani2016analysis}. We repeat experiments on 4 random seeds.

We count flops and parameters using the thop package (for package version numbers see automated logs in the online appendix linked above): \url{https://github.com/Lyken17/pytorch-OpCounter}

Code for running the experiment is available at:\\ \url{https://github.com/Breakend/ClimateChangeFromMachineLearningResearch/blob/master/paper_specific/run_inference.py}

An online appendix showing all per-experiment details can be seen here:\\ \url{https://breakend.github.io/ClimateChangeFromMachineLearningResearch/measuring_and_mitigating_energy_and_carbon_footprints_in_machine_learning/}

\begin{figure}[!htbp]
    \centering
    \includegraphics[width=.49\textwidth]{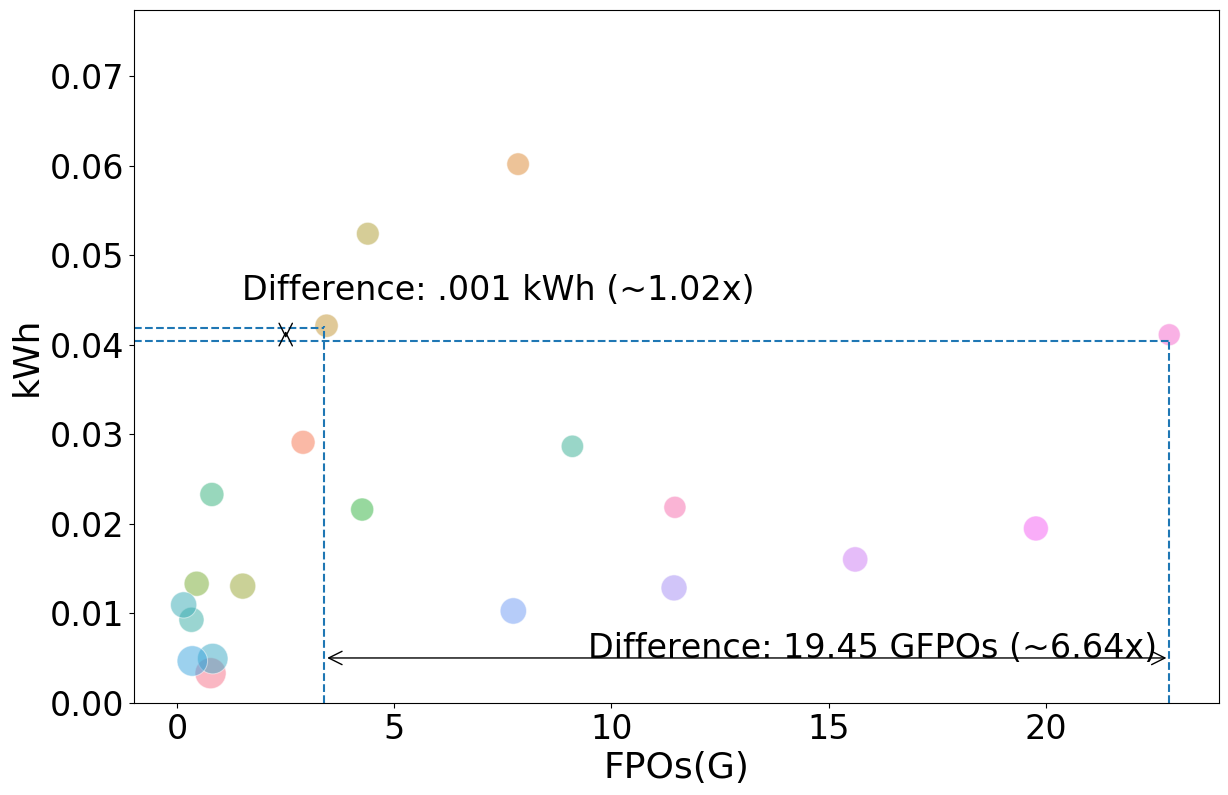}
    \includegraphics[width=.49\textwidth]{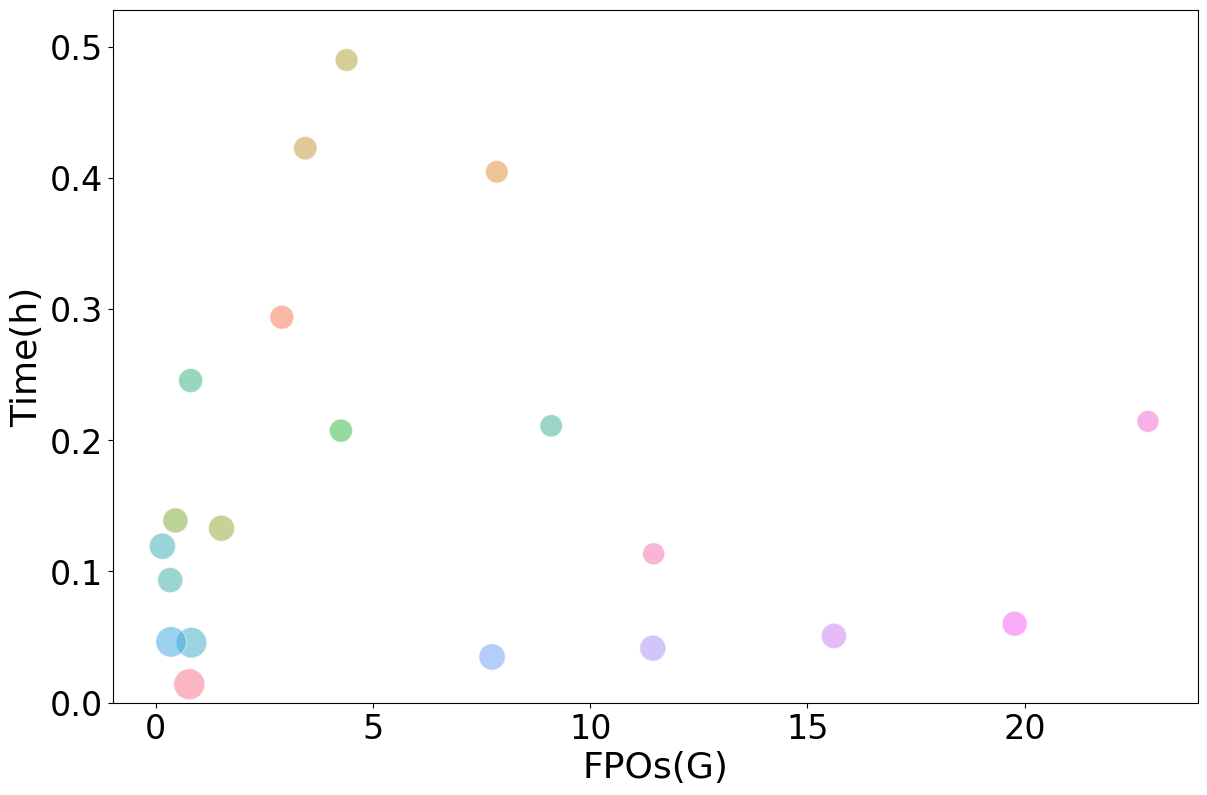}
    \includegraphics[width=\textwidth]{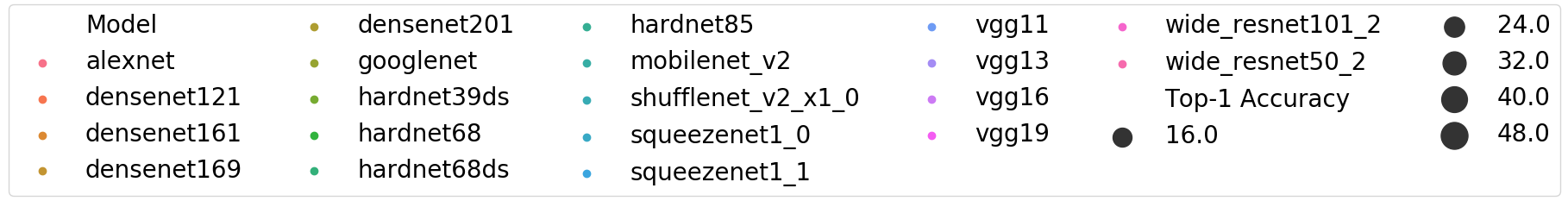}
    \caption{We seek to investigate the connection between FPOs, energy usage, and experiment time, similarly to \citet{canziani2016analysis}. We run 50,000 rounds of inference on a single image through pre-trained image classification models available through PyTorch Hub (see \url{https://pytorch.org/hub})---namely~\citep{krizhevsky2012imagenet,huang2017densely,43022,chao2019hardnet,sandler2018mobilenetv2,zhang2018shufflenet,iandola2016squeezenet,simonyan2014very,zagoruyko2016wide}. We record experiment time and the kWh of energy used to run the experiments and repeat experiments 4 times, averaging results. We find that FPOs are not strongly correlated with energy consumption ($R^2=0.083$, Pearson $0.289$) nor with time ($R^2 = 0.005$, Pearson $-0.074$). Number of parameters (plotted in Appendix) similarly have no strong correlation with energy consumption ($R^2 = 0.002$, Pearson $-0.048$), nor time ($R^2 = 0.14$ , Pearson $-.373$). We note, however, that \emph{within an architecture} correlations are much stronger. For example, only considering different versions of VGG, FPOs are strongly correlated with energy ($R^2=.999$, Pearson $1.0$) and time ($R^2=.998$, Pearson $.999$). See Appendix for experiment details, code, and data links. Our runtime results likely differ from \citet{canziani2016analysis} due to the architectural differences in the model sets we use.}
    \label{fig:imagenet2}
\end{figure}

\begin{figure}[!htbp]
    \centering
    \includegraphics[width=.49\textwidth]{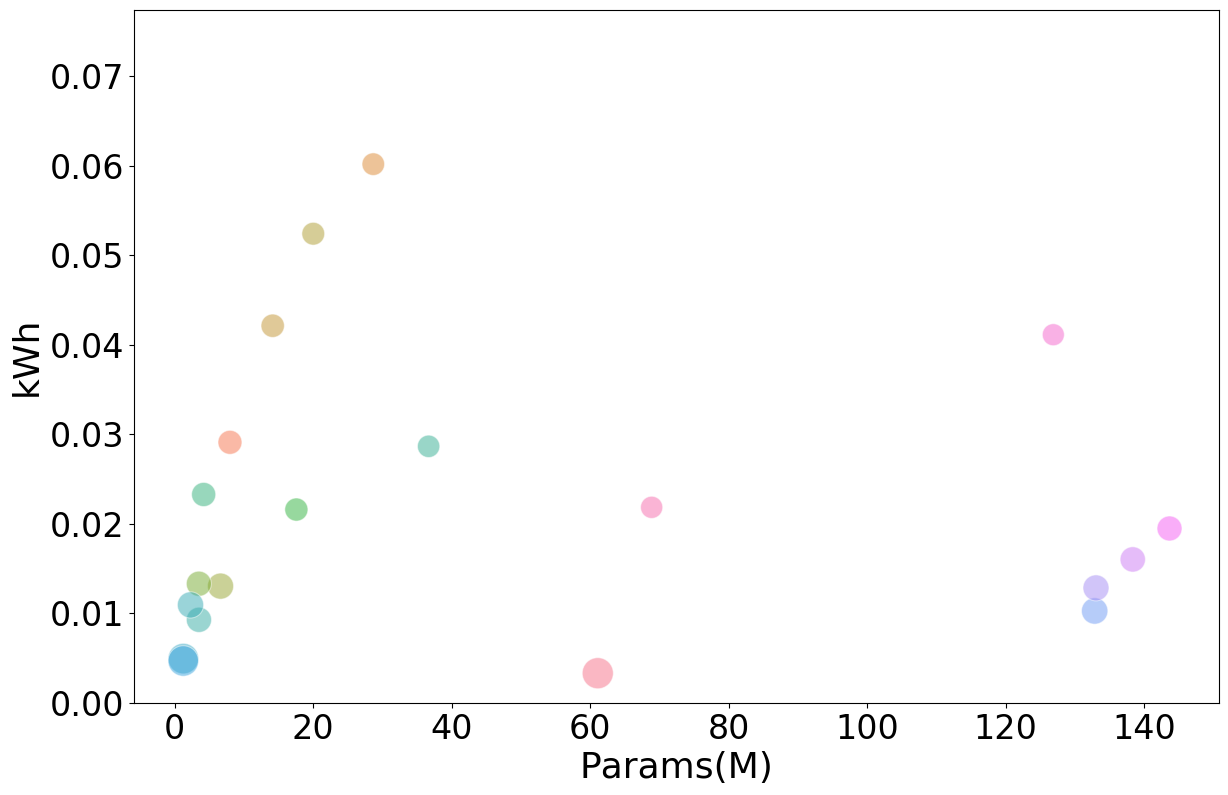}
    \includegraphics[width=.49\textwidth]{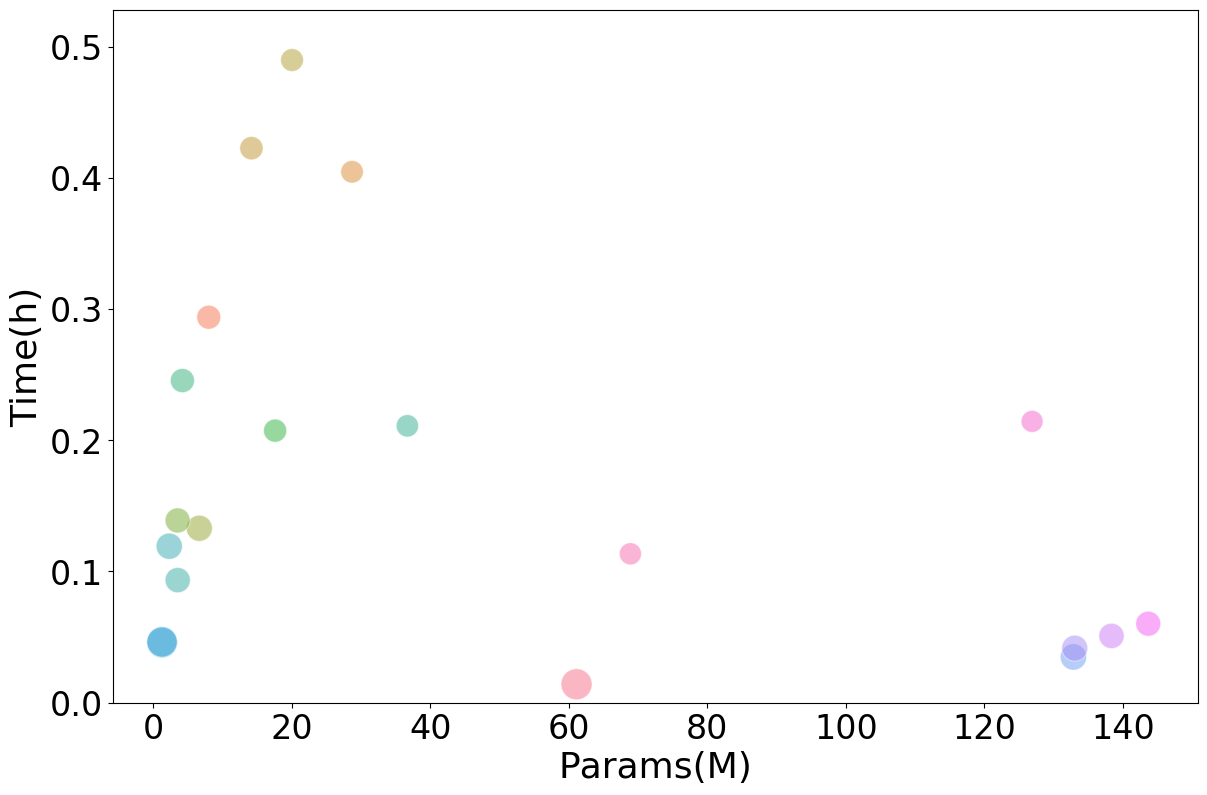}
    \caption{The same experiments as in Figure~\ref{fig:imagenet}, plotting parameters as the varying factor instead. See Figure~\ref{fig:imagenet} for correlation values.}
    \label{fig:parameter_imagenet_plots}
\end{figure}

The plot of FPOs versus runtime can be seen in Figure~\ref{fig:imagenet2} and plots against number of parameters can be seen in Figure~\ref{fig:parameter_imagenet_plots}. Number of parameters similarly have no strong correlation with energy consumption ($R^2 = 0.002$, Pearson $-0.048$), nor time ($R^2 = 0.14$ , Pearson $-.373$). We note that our runtime results likely differ from \citet{canziani2016analysis} due to the architectural differences in the model sets we use.

For parameter plots, see Figure~\ref{fig:parameter_imagenet_plots}, for extended time and energy Figures, see Figure~\ref{fig:imagenet2}.

\section{Estimation Methods}
\label{app:fig2}

We use our PPO Pong experiment (see Appendix~\ref{app:rl} for more details) as the experiment under comparison. For carbon emission estimates, we use three estimation methods: realtime emissions data for California (collected by our framework from \url{caiso.org}) times the power usage at that time integrated over the length of the experiment; multiplying total energy usage recorded by our method by the California average carbon intensity; multiplying total energy usage recorded by our method by the EPA US average carbon intensity~\citep{strubell2019energy}. For energy estimates, we use: (1) the experiment time multiplied by the number of GPUs, a utilization factor of 1/3 or 1, and the Thermal Design Power (TDP)---which can be thought of as the maximum Watt draw---of the GPU~\citep{aiandcompute}; (2) the measured GPU-hrs of our tool multiplied by the TDP; a rough calculation of PFLOPs-hr (following the methodology of \citep{aiandcompute} by the PFLOPs/TDP of the GPU; (3) our tool's accounting method which tracks energy from GPU readings, accounts for CPU time/energy, and measures utilization in realtime.

\section{Reinforcement Learning}
\label{app:rl}

We investigate the energy efficiency of four baseline RL algorithms: PPO \citep{stable-baselines,schulman2017proximal}, A2C~\citep{stable-baselines,mnih2016asynchronous}, A2C with VTraces~\citep{espeholt2018impala,dalton2019gpuaccelerated}, and DQN~\citep{stable-baselines,mnih2016asynchronous}. 
We evaluate on PongNoFrameskip-v4 (left) and BreakoutNoFrameskip-v4 (right), two common evaluation environments included in OpenAI Gym~\citep{bellemare2013arcade,1606.01540,mnih2013playing}. 

We train for only 5M timesteps, less than prior work, to encourage energy efficiency~\citep{mnih2016asynchronous,mnih2013playing}.
We use default settings from code provided in stable-baselines \citep{stable-baselines} and cule \citep{dalton2019gpuaccelerated}, we only modify evaluation code slightly.
Modifications can be found here:

\begin{itemize}
    \item \url{https://github.com/Breakend/rl-baselines-zoo-1} (for stable-baselines modifications)
    \item \url{https://github.com/Breakend/cule} (for cule modifications)
\end{itemize}

Since we compare both on-policy and off-policy methods, for fairness all evaluation is based on 25 separate rollouts completed every 250k timesteps. This is to ensure parity across algorithms. We execute these in parallel together as seen in the cule code: \url{https://github.com/Breakend/cule/blob/master/examples/a2c/test.py}. 

\begin{figure}[H]
    \centering
    \includegraphics[width=.49\textwidth]{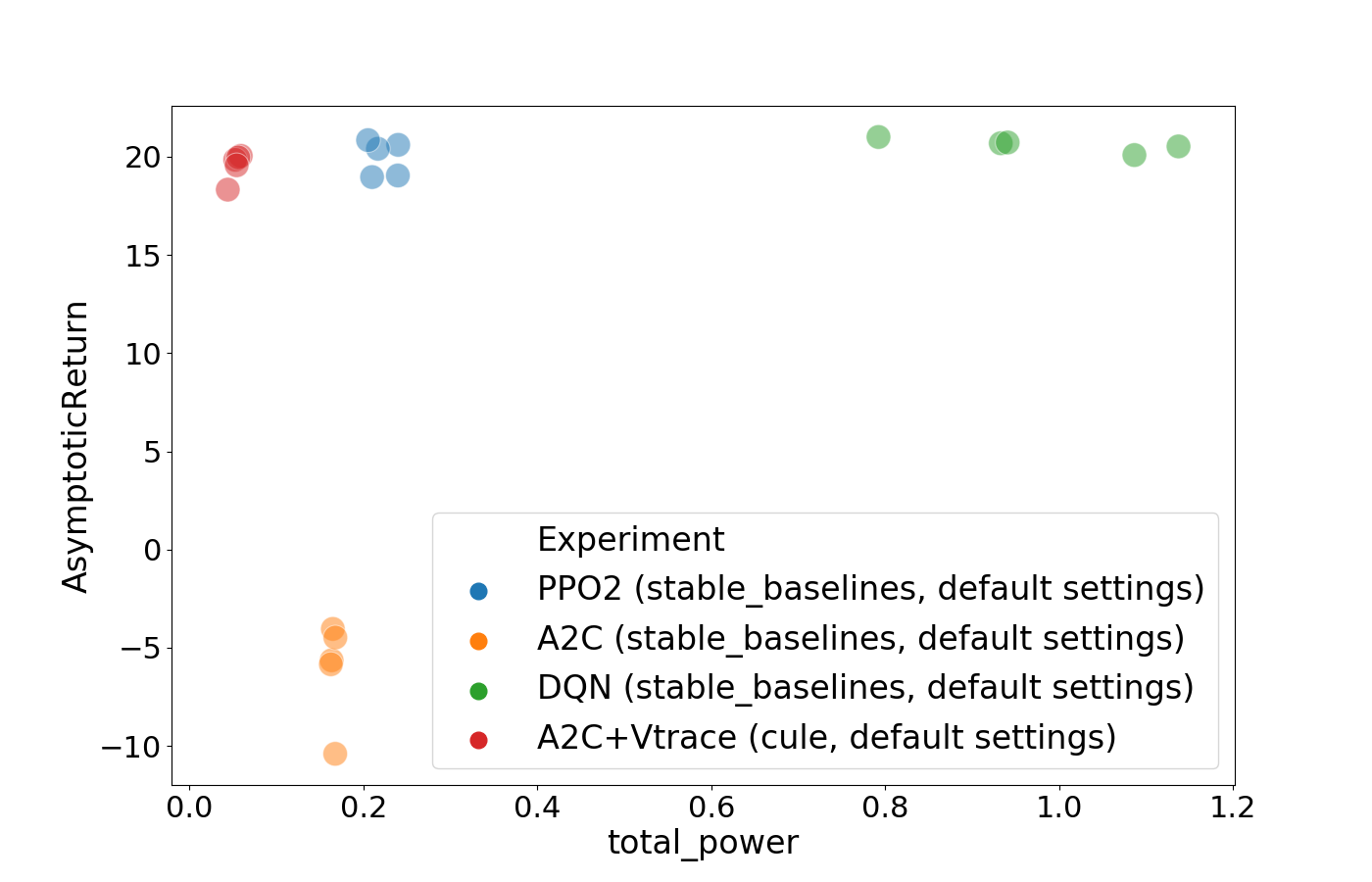}
    \includegraphics[width=.49\textwidth]{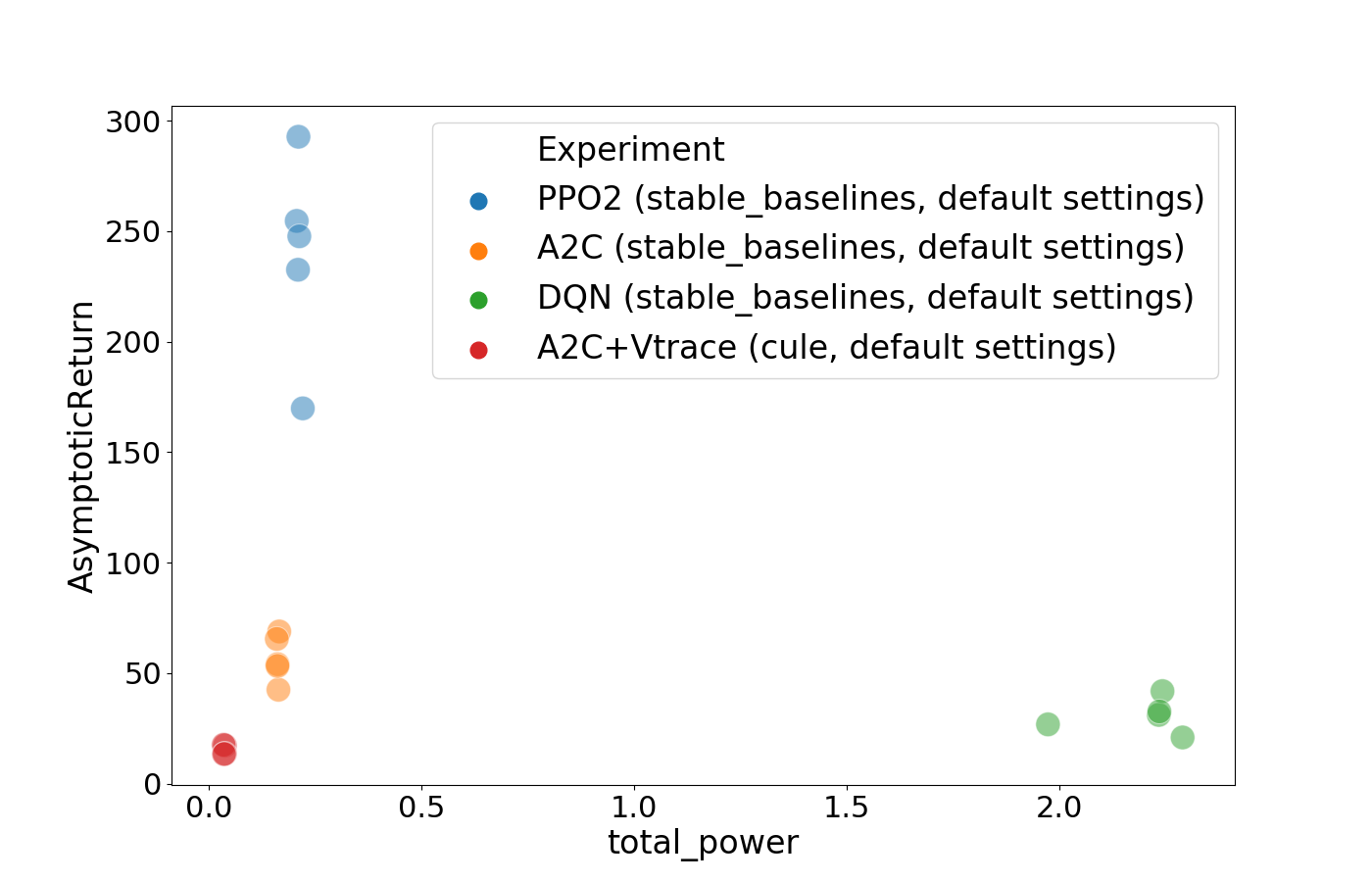}
    \caption{Pong (left) and Breakout (right) asymptotic return.}
    \label{fig:rl_asymptotic}
\end{figure}

\begin{figure}[H]
    \centering
    \includegraphics[width=.49\textwidth]{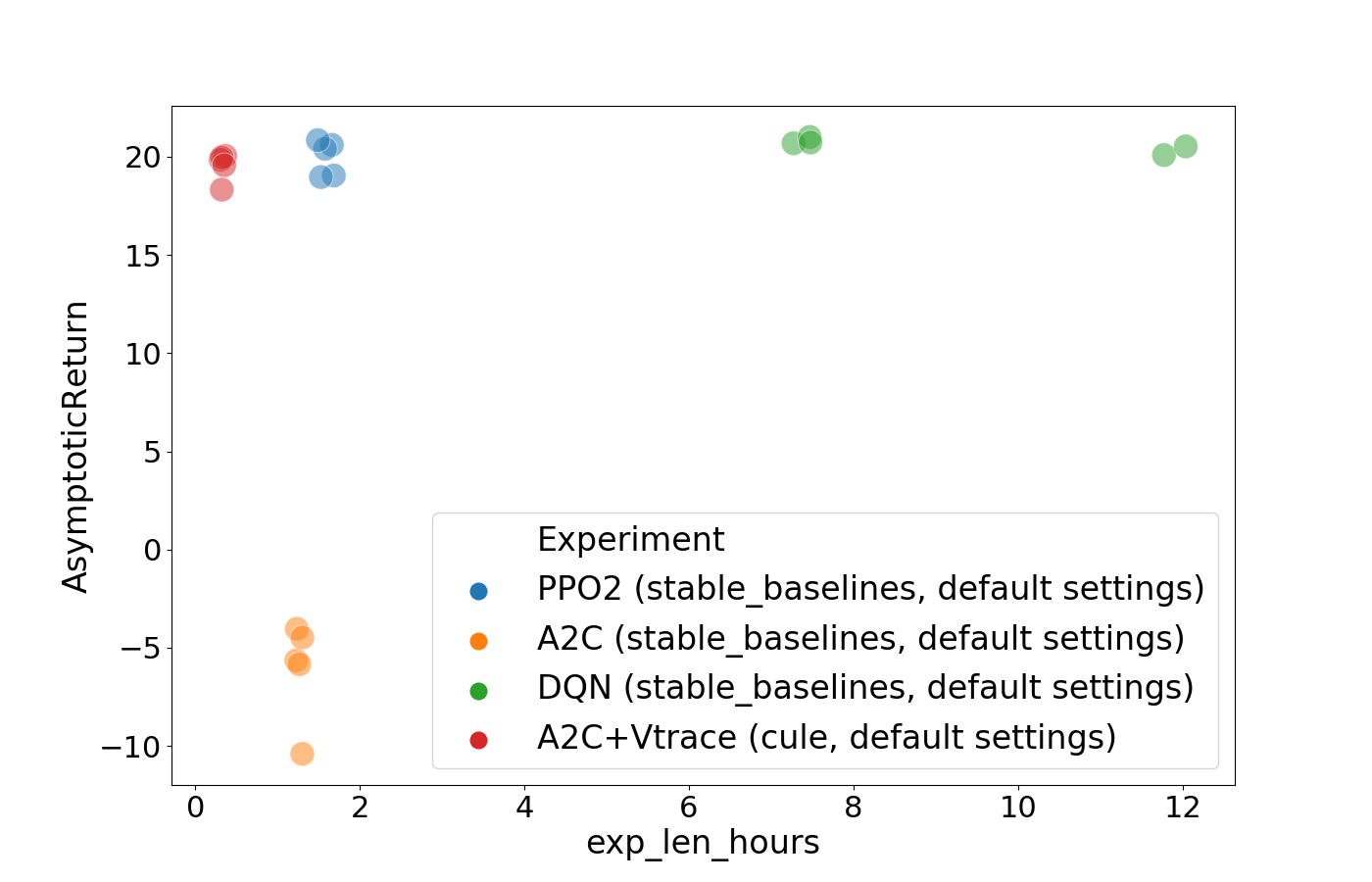}
    \includegraphics[width=.49\textwidth]{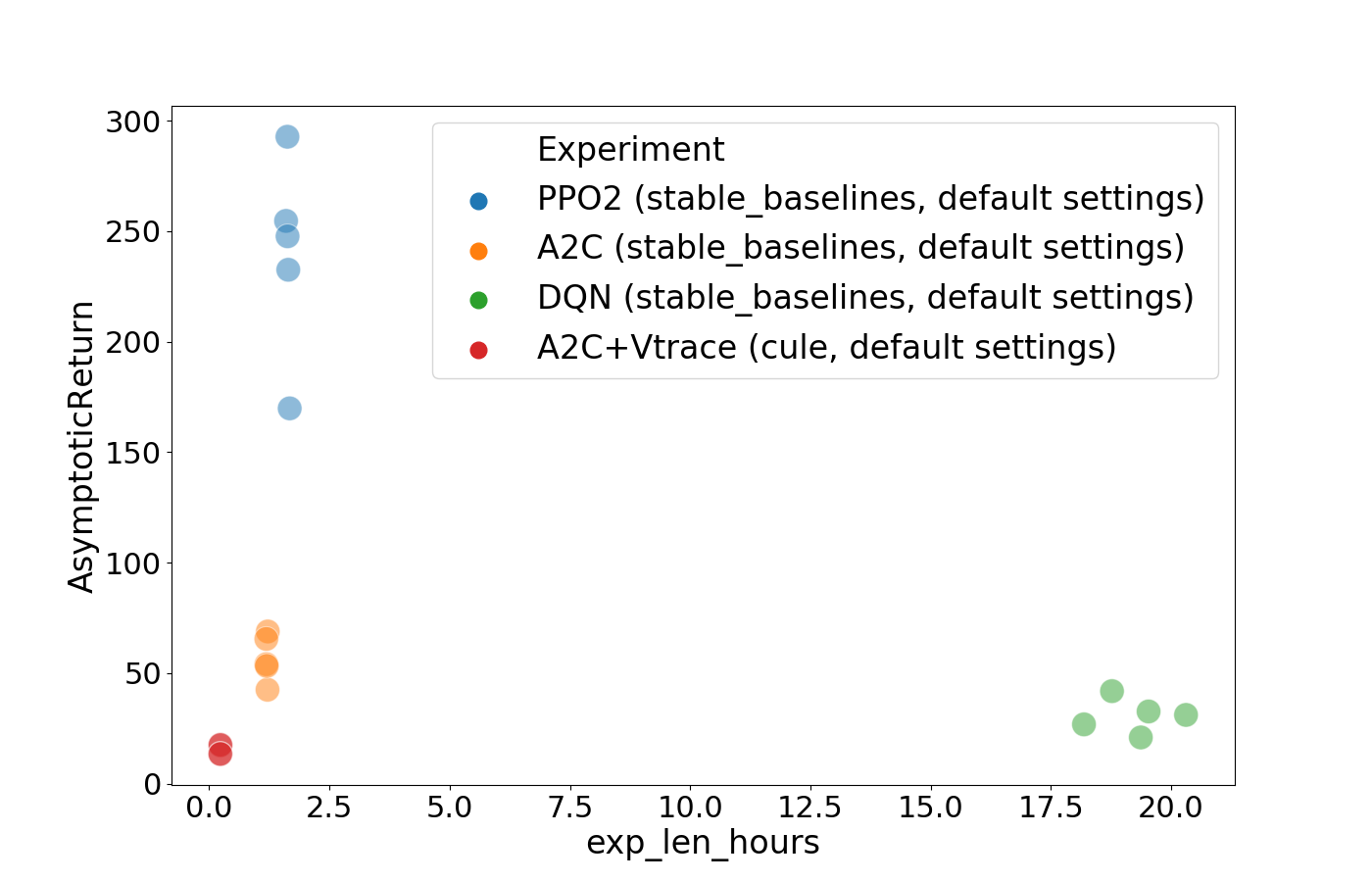}
    \caption{Pong (left) and Breakout (right) as a function of experiment length and asymptotic return.}
    \label{fig:rl_time_asymptotic}
\end{figure}

\begin{figure}[H]
    \centering
    \includegraphics[width=.49\textwidth]{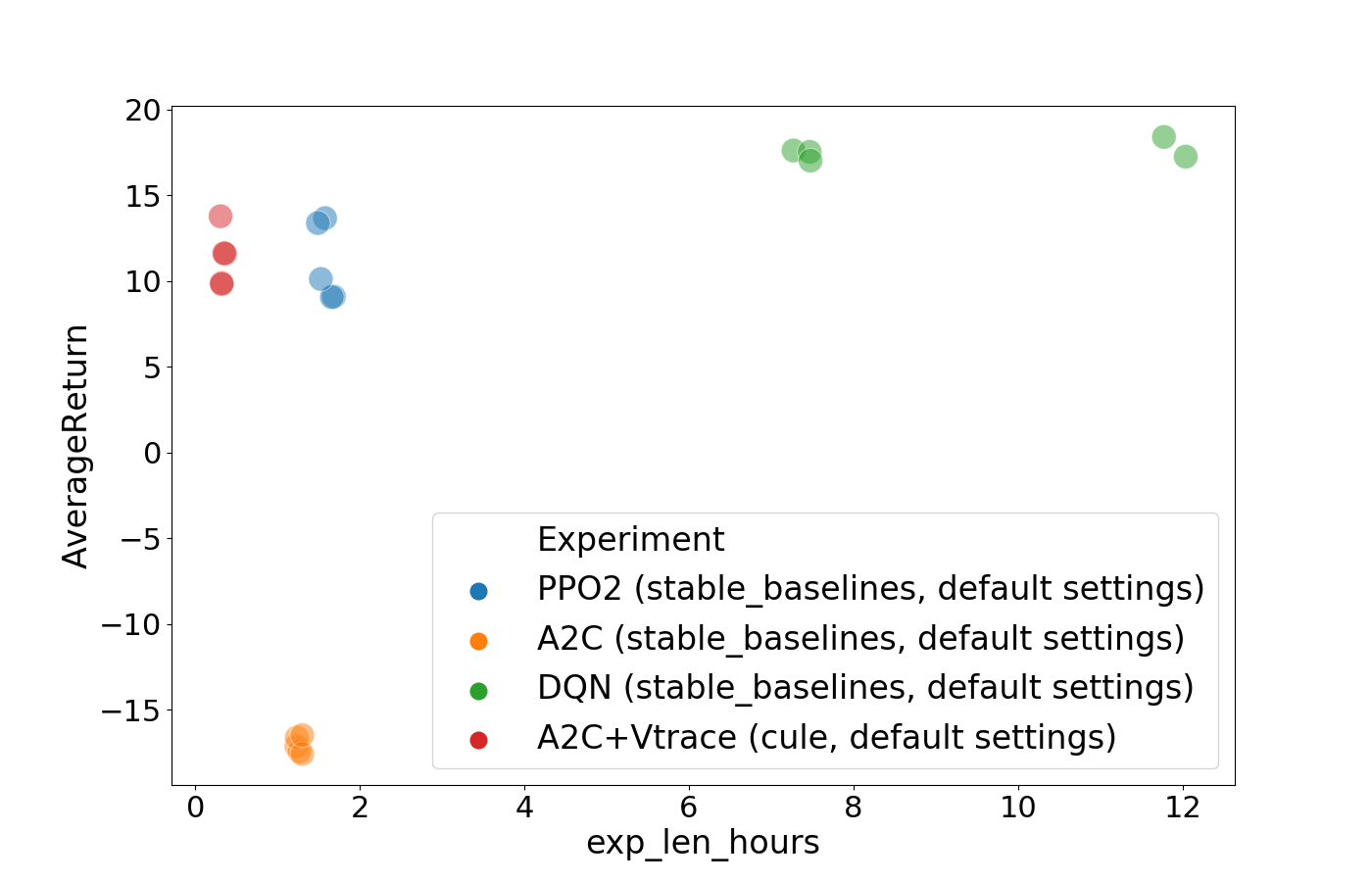}
    \includegraphics[width=.49\textwidth]{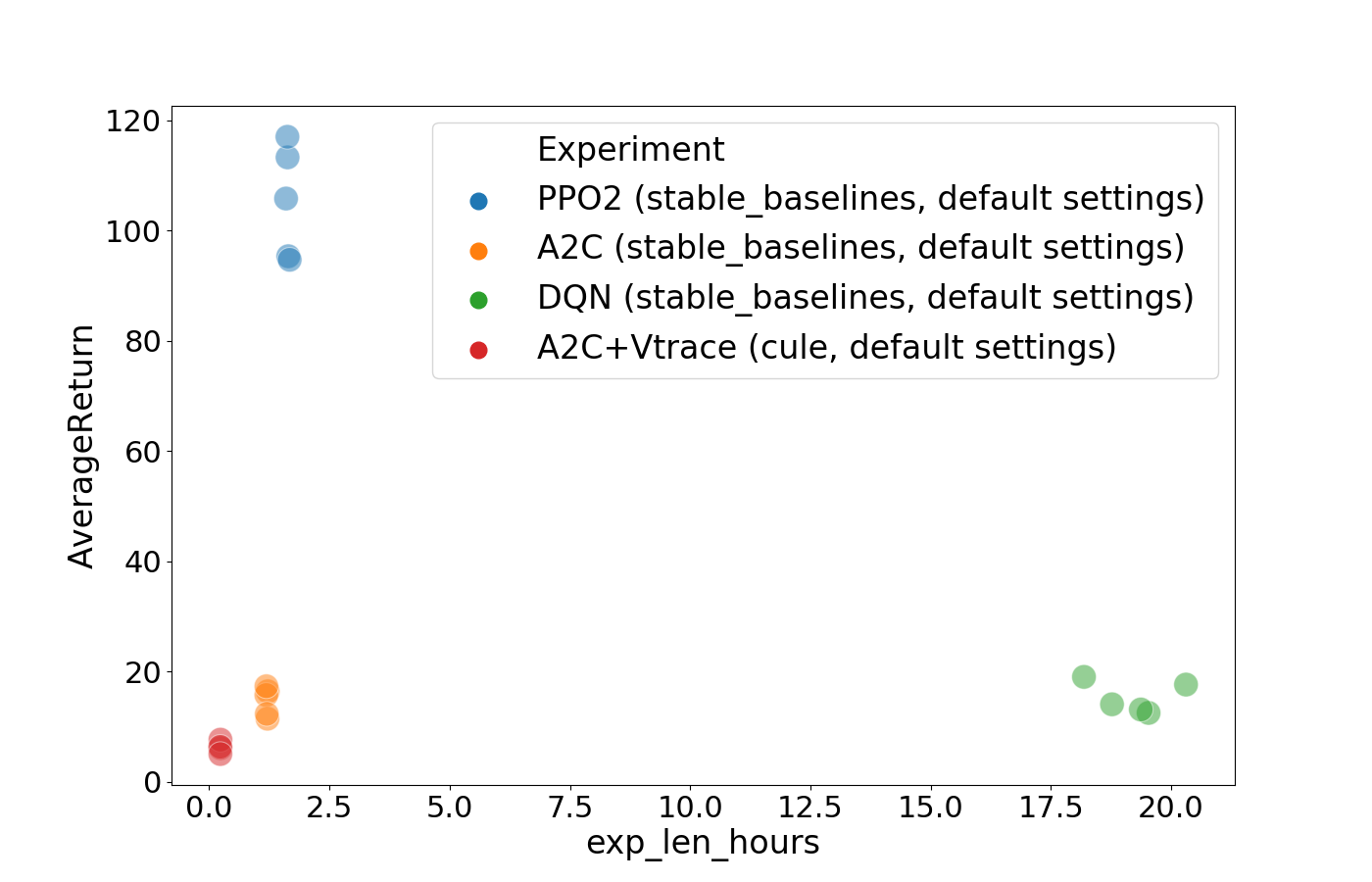}
    \caption{Pong (left) and Breakout (right) as a function of experiment length and average return.}
    \label{fig:rl_time_average}
\end{figure}

While average return across all evaluation episodes (e.g., averaging together the step at 250k timesteps and every evaluation step until 5M timesteps) can be seen in the main text, the asymptotic return (for the final round of evaluation episodes) can be seen Figure~\ref{fig:rl_asymptotic}. Plots comparing experiment runtime to asymptotic and average returns (respectively) can be seen in Figure~\ref{fig:rl_time_asymptotic} and Figure~\ref{fig:rl_time_average}.

Our online leaderboard can be seen at: \url{https://breakend.github.io/RL-Energy-Leaderboard/reinforcement_learning_energy_leaderboard/index.html}

We note that while DQN underperforms as compared to PPO here, better hyperparameters may be found such that DQN is the more energy efficient algorithm. Moreover, we only use the 5M samples regime, whereas prior work has used 10M or more samples for training, so DQN results seen here would correspond to earlier points in training in other papers.

\section{Possible Sources of Error, Limitations, and Overheads} 
\label{sec:sourcesoferror}
In Sections~\ref{sec:bad_estimates} and~\ref{sec:flops}, we compared different methods for estimating energy and carbon emissions including extrapolating from FPOs. However, we note that our own framework is not perfect. For transparency, we highlight several such sources here, but we note that utilizing more information---as we do here---is by definition superior to approximations which rely on less accurate assumptions (see Section~\ref{sec:bad_estimates}). 

First, we rely on downstream hardware APIs which themselves have errors. Several works have sought to evaluate the accuracy of RAPL---see for example \citet{desrochers2016validation} and \citet{kavanagh2019rapid}---and Nvidia's power profiling tool---see for example, \citet{sen2018quality} and \citet{arafa2020verified}. Errors highly depend on the specific chipset and even the workload, so we refer the reader to these other works for techniques in assessing exact errors. Nvidia's documentation, however, states that the power reading ``is  accurate  to  within  +/- 5 watts.''\footnote{\href{https://developer.download.nvidia.com/compute/DCGM/docs/nvidia-smi-367.38.pdf}{https://developer.download.nvidia.com/compute/DCGM/docs/nvidia-smi-367.38.pdf}}

Second, we rely on a polling mechanism due to the constraints of these downstream APIs (for GPUs typically only power is provided, rather than an energy counter). In particularly short jobs or highly erratic workloads, the tool may poll at a time that is not representative of the full workload, estimating energy usage from an outlier power sample. Our assumption is that workloads are fairly consistent and long enough that such variability will average out. In the event that comparisons of energy readings across models are needed, we encourage users to report standard errors across several runs (with $n$ appropriate for the experiment setting). Furthermore, because we record many auxiliary data sources (such as CPU frequency), more accurate estimates can further be conducted via mixed effects models to control for sources of variation and noise in energy readings. For an example of how such an analysis would work, see for example \citet{boquet2019decovac}, which compare machine learning algorithms controlling for hyperparameter choice and randomness.

Third, for cloud regions, we do not have access to the exact carbon intensities or PUEs. For example, if a cloud provider has a direct connection to a clean energy power plant for 100\% of its energy, we have no way of accessing this information and using it in our tool. We encourage companies to report this information per cloud region so that this may be more accurate. In the case of indirect carbon offsetting, we do not consider this to be an inaccuracy---see discussion in Appendix~\ref{app:carbon_discussion}. Moreover, we rely on IP address information and hand-gathered energy grid information to estimate the energy grid. Either of these may incur errors. Since we report this information and allow users to override grid regions in calculations, these may be corrected by users. We also may not be able to access particular drivers needed on every cloud instance. As such, support may depend on the cloud machine image being used and the drivers available on that image. Generally, if Intel's RAPL is available or PowerGadget can be installed---and nvidia-smi is available---then the system should be compatible.

Regarding overheads to adding a separate process gathering these metrics, the cost should be generally fairly low. There are some startup and shutdown costs associated with adding the tool, so for short-running scripts the absolute percentage of overhead may be he higher. Additionally, if computational capacity of a chipset is maximally used due to the main process, there may be some added cost for thread switching to gather metrics. However, assuming that a core is preserved for the impact tracker there should be minimal overhead. Note, for the sake of reproducibility we also record disk read/write speeds, but this can be turned off if the disk is particularly slow or there is too much disk I/O for the user's liking. While workload overhead can vary depending on the machine and workload, we found that in a small experiment of 200 epochs of regression for a one hidden layer neural network, runtime overheads were less than 1\%. For 500 epochs, the overhead was around .5\% (indicating that startup/shutdown are the most intensive). This experiment was run on a CPU-only Mac OS machine with a 2.7 GHz Quad-Core Intel Core i7 and 16 GB 2133 MHz LPDDR3.

Supporting every driver and hardware combination is difficult. We note that most of the aforementioned metrics are only supported on Linux systems and we were only able to test hardware combinations available to us. Mac OS support is limited to machines that have Intel's Power Gadget\footnote{\href{https://software.intel.com/content/www/us/en/develop/articles/intel-power-gadget.html}{https://software.intel.com/content/www/us/en/develop/articles/intel-power-gadget.html}} installed and to CPU-only recordings. We hope that future users will help identify missing capabilities and expand the framework for new use-cases and machines. We also note that the tool is limited by driver support in cases that we cannot work around (see Section~\ref{sec:drivers}). 

Finally, we note that we only record CPU, GPU, and DRAM power draw. We do not record disk I/O energy usage, power conversion and voltage regulator overhead.  As such, we can expect there to be missing components that contribute to energy that we do not record here. However, we expect that the PUE re-scaling will correct for some of these missing components to some extent.

\section{Comparing Models}
\label{app:comparisons}
We note that it may be tempting to use carbon emissions as a comparative tool: model A is less carbon intensive that model B. However, unless the carbon intensity used for either model is held constant, this comparison cannot be done. In particular, our tool should not be used to compare carbon emissions between models without overriding the carbon intensity used as we sometimes use real-time values. If two models are compared, as in Section~\ref{sec:rl}, multiple runs on comparable machines should be used. In the event that a robust conclusion is to be made (e.g., Algorithm A is more energy efficient than Algorithm B), additional metrics regarding workload that we record can be utilized to run a mixed-effects regression analysis. Such an analysis would ensure that there aren't confounding factors jeopardizing the conclusion.

\end{document}